\documentclass[11pt]{article}
\pdfoutput=1

\usepackage{a4wide}
\usepackage{amsmath}
\usepackage{amssymb}
\usepackage{xfrac}
\usepackage{nicefrac}

\usepackage{mathrsfs}
\usepackage{mathpazo}

\usepackage{tikz}
\usetikzlibrary{matrix}

\tikzset{
  pretableaumatrix/.style={
    ampersand replacement=\&,
    matrix of math nodes,
    outer sep=1mm,
    inner sep=0mm,
    anchor=center,
    row sep={between borders,-\pgflinewidth},
    column sep={between borders,-\pgflinewidth},
    dottedentry/.style={densely dotted},
    dashedentry/.style={densely dashed},
    spaceentry/.style={draw=none,execute at begin node=\null},
  },
  pretableaunode/.style={
    font=\small,
    draw=gray,
    sharp corners,
    rectangle,
    anchor=base,
    text height=3.75mm,
    text depth=1.25mm,
    minimum height=5mm,
    minimum width=5mm,
    inner sep=0mm,
    outer sep=0mm,
    doublewidth/.style={minimum width=10mm},
    footnotesize/.style={font=\footnotesize},
    scriptsize/.style={font=\scriptsize},
  },
  tableaumatrix/.style={
    pretableaumatrix,
    every node/.append style={
      pretableaunode,
    },
  },
  medtableaumatrix/.style={
    pretableaumatrix,
    every node/.append style={
      pretableaunode,
      font=\footnotesize,
      text height=2.75mm,
      text depth=.75mm,
      minimum height=3.5mm,
      minimum width=3.5mm
    },
  },
  smalltableaumatrix/.style={
    pretableaumatrix,
    every node/.append style={
      pretableaunode,
      font=\scriptsize,
      text height=1.85mm,
      text depth=.15mm,
      minimum height=2.5mm,
      minimum width=2.5mm,
    },
  },
  tinytableaumatrix/.style={
    pretableaumatrix,
    every node/.append style={
      pretableaunode,
      font=\tiny,
      text height=1.25mm,
      text depth=.15mm,
      minimum height=1.75mm,
      minimum width=1.75mm
    },
  },
  tableau/.style={
    baseline=-1.25mm,
    every matrix/.style={tableaumatrix},
  },
  medtableau/.style={
    baseline=-1.25mm,
    every matrix/.style={medtableaumatrix},
  },
  smalltableau/.style={
    baseline=-1.25mm,
    every matrix/.style={smalltableaumatrix},
  },
  preshapetableaumatrix/.style={
    pretableaumatrix,
    execute at end cell={\strut},
    every node/.append style={
      draw=black,
      anchor=base,
      inner sep=0mm,
      outer sep=0mm,
    },
    shadedentry/.style={fill=gray},
    darkshadedentry/.style={fill=darkgray},
  },
  medshapetableaumatrix/.style={
    preshapetableaumatrix,
    every node/.append style={
      text height=2.75mm,
      text depth=.75mm,
      minimum height=3.5mm,
      minimum width=3.5mm
    },
  },
  shapetableaumatrix/.style={
    ampersand replacement=\&,
    matrix of math nodes,
    outer sep=0mm,
    inner sep=0mm,
    anchor=base,
    row sep={between borders,-\pgflinewidth},
    column sep={between borders,-\pgflinewidth},
    execute at begin cell={\strut},
    every node/.append style={draw,anchor=base,text height=1mm,text depth=.5mm,minimum size=1.5mm,inner sep=0mm,outer sep=0mm},
  },
  shapetableau/.style={
    every matrix/.style={shapetableaumatrix},
  },
  topalign/.style={
    every matrix/.append style={name=maintableau,anchor=maintableau-1-1.base},
    baseline,
  },
}

\RequirePackage{amsmath,amssymb,amsxtra,amsthm}
\usepackage{nicefrac, xfrac}

\RequirePackage{mathrsfs}

\RequirePackage{setspace}
\RequirePackage{array}
\RequirePackage{booktabs}

\RequirePackage{colortbl}
\RequirePackage{xcolor}
\colorlet{titlerowcolor}{gray!15}

\usepackage{cite}

\definecolor{blue3}{RGB}{31,119,180}
\definecolor{red3}{RGB}{214,39,40}
\definecolor{orange3}{RGB}{255,127,14}
\definecolor{green3}{RGB}{44,160,44}

\usepackage{colortbl}
\definecolor{lightgreen}{cmyk}{0.2, 0, 0.2, 0.2}
\definecolor{lightgray}{cmyk}{0.1,0.2,0,0.1}
\definecolor{lightgray2}{cmyk}{0.1,0.1,0,0.1}

\usepackage{hyperref}
\hypersetup{colorlinks=true,linkcolor=red3,citecolor=green3,urlcolor=teal,pdfencoding=auto}

\newcommand{\Jtheta}[4]{ \theta \big[  \substack{\scriptstyle{#1} \\ \scriptstyle{#2}} \big] 	 ( #3 | #4  ) }

\newcommand{\Jacobitheta}[2]{ \theta \big[ \substack{\scriptstyle{#1} \\ \scriptstyle{#2}} \big] 	 }

\newcommand{\Jacobibartheta}[2]{ \bar \theta \big[  \substack{\scriptstyle{#1} \\ \scriptstyle{#2}}  \big] 	 }

\numberwithin{equation}{section}
\numberwithin{table}{section}
\numberwithin{figure}{section}

\author{
  \begin{minipage}{1.00\linewidth}
    \vspace{1cm}
    \begin{center}
      \begin{small}
        \textbf{ Giorgio Leone}
        \\ \vspace{1cm}
        {\em Dipartimento di Fisica e Astronomia “Galileo Galilei”, Università degli Studi di Padova}
        \\
        {\em INFN Sezione di Padova, Via F. Marzolo 8, 35131 Padova, Italy}
     \end{small}
    \end{center}
    \vspace{1cm}
  \end{minipage}
}

\date{}

\title{\vspace{3cm}
  \begin{huge} \textbf{Non-supersymmetric strings on AdS\texorpdfstring{$_3$}{3}: a world-sheet perspective} 	
  \end{huge}
  \\ \vspace{.7cm}
}

\begin{document}

\begin{titlepage}
  \maketitle
  \thispagestyle{empty}

  \vspace{-14cm}
  \begin{flushright}
   \end{flushright}

  \vspace{11cm}

  \begin{center}
    \textsc{Abstract}\\
  \end{center}

\noindent We explore the quantisation of the tachyonic type 0B superstring and the non-tachyonic $\text{Spin}(16) \times \text{Spin}(16) \rtimes \mathbb{Z}_2$ heterotic string on AdS$_3 \times S^3 \times T^4$ and  AdS$_3 \times S^3 \times S^3 \times S^1$ backgrounds. Adapting the analysis for the supersymmetric and bosonic string theories to these set-ups, we provide a world-sheet description for a generic level of the $\text{SL}(2,\mathbb{R})$ WZW model, and we read the spectrum through the associated partition functions. Focusing on the low-energy theory, we show that the $\text{Spin}(16) \times \text{Spin}(16) \rtimes \mathbb{Z}_2$ heterotic string on both backgrounds accommodates non-trivial Wilson lines that are responsible for the appearance of tachyonic regions in the classical moduli space, hence jeopardising the stability of the vacuum. We show this with a concrete example on the AdS$_3 \times S^3 \times S^3 \times S^1$ space and provide general formulas for a systematic analysis of the classical moduli space.

\vfill

{\small
\begin{itemize}
\item[E-mail:]  {\tt giorgio.leone@unipd.it}

\end{itemize}
}

\end{titlepage}

\setstretch{1.1}


{		\hypersetup{linkcolor=black}
		\tableofcontents	}

\newpage

\section{Introduction}

Non-supersymmetric string theories are plagued by instabilities. In the vast majority of cases, such instabilities may already appear classically by the presence of tachyons in the tree-level spectrum. Although a proper algorithm ensuring the absence of tachyons for a given string construction is missing, few exceptions are known to be tachyon-free. In ten dimensions these amount to only three theories: the Sugimoto model \cite{Sugimoto:1999tx}, the type 0'B superstring or the {\em Sagnotti} model \cite{Sagnotti:1995ga,Sagnotti:1996qj} and the $\text{Spin}(16) \times \text{Spin}(16) \rtimes \mathbb{Z}_2$ heterotic string \cite{Alvarez-Gaume:1986ghj,Dixon:1986iz}.
However, even though these theories do not have any tachyonic instabilities, divergences appear in higher genus amplitudes. Indeed, they occur whenever the Riemann surface factorises into two non-vanishing sub-diagrams connected by an infinitely long cylinder in which massless scalars at zero momentum propagate. Hence, these divergences have a natural IR interpretation and, as such, are not pathological for the consistency of the theory itself. In particular, it is known (at least in principle) how to take care of this issue: one can compensate these infinities by introducing suitable counter-terms in the non-linear sigma model \cite{Lovelace:1986kr,Fischler:1986ci, Fischler:1986tb}. These counter-terms, on the other hand, modify the structure of the $\beta$-function or, equivalently, of the effective action via runaway potentials. Therefore, when higher-genus corrections are taken into account, the vanilla tree-level vacuum we have started with--for instance, Minkowski--does not make the loop-corrected $\beta$ function vanish, and hence it is not a valid vacuum anymore. 
A full-fledged stringy analysis is generally speaking hard to put forward since, for most of these backgrounds, a proper world-sheet description is unknown. One can only rely on a low-energy effective field theory analysis which has been nonetheless extremely useful in understanding many features of these theories \cite{Dudas:2000ff,DeWolfe:2001nz,Gubser:2001zr,Dudas:2002dg,Dudas:2010gi,Mourad:2016xbk,Basile:2018irz,Mourad:2021qwf,Mourad:2021roa,Mourad:2022loy,Antonelli:2019nar,Basile:2020mpt,Basile:2022vft,Basile:2021krk,Basile:2022ypo,Raucci:2022bjw,Raucci:2022jgw,Mourad:2024dur,Mourad:2024mpg,Robbins:2025wlm,Raucci:2025bev}. 

Recently, it was argued that the $\text{Spin}(16) \times \text{Spin}(16) \rtimes \mathbb{Z}_2$ heterotic string admits AdS$_3 \times S^3 \times S^3 \times S^1$ and AdS$_3 \times S^3 \times T^4$ backgrounds as solutions of the cancellation of the one-loop corrected Weyl anomaly \cite{Baykara:2022cwj,Robbins:2025wlm}. In particular, two classes have been identified: backgrounds admitting a well-defined AdS$_3$ target space at tree-level, and backgrounds which only exist if quantum corrections are taken into account, hence dubbed in the following {\em purely quantum}. For the latter, the AdS$_3$ curvature contribution to the $\beta$-function is compensated by the presence of the runaway potential and hence no AdS$_3$ theory can be consistently defined on the sphere. These backgrounds, therefore, go beyond the domain of application of the standard CFT techniques, and, to date, no world-sheet description is available to study them. For the former, the theory can be consistently defined on the sphere, for which the AdS$_3$ curvature contribution to the $\beta$-function is compensated by flux units of the Kalb-Ramond field, allowing for a fully-fledged world-sheet description. These backgrounds have been extensively studied in the literature for the bosonic string and supersymmetric superstrings \cite{Giveon:1998ns,Maldacena:2000hw,Maldacena:2000kv,Maldacena:2001km,Israel:2003ry,Giribet:2007wp,Ferreira:2017pgt,Eberhardt:2017fsi,Eberhardt:2021vsx}, which lead to the identification of the holographic dual CFTs both in the tensionless limit \cite{Gaberdiel:2018rqv,Giribet:2018ada,Eberhardt:2018ouy,Eberhardt:2019ywk,Gaberdiel:2024dva} and at finite tension \cite{Eberhardt:2019qcl,Eberhardt:2021vsx}. 
Hence, although in the following we will not be interested in holographic applications, we can adapt the technology developed in these recent years to the case of strings without space-time supersymmetry, including the $\text{Spin}(16) \times \text{Spin}(16) \rtimes \mathbb{Z}_2$ heterotic string at hand. For such a theory the analysis was performed in the limit in which the AdS$_3$ background reproduces flat Minkowski \cite{Baykara:2022cwj,Fraiman:2023cpa}. This approximation holds for large values of the curvature radius, which, however,  does not correspond to the most general scenario. 
Moreover, it is natural to ask whether introducing marginal operators corresponding to Wilson lines could be a threat to the stability of these theories, as it happens in flat space with the same internal manifolds.
In this paper, we thus aim to fill these gaps by providing a world-sheet description for the $\text{Spin}(16) \times \text{Spin}(16) \rtimes \mathbb{Z}_2$ heterotic string on such AdS$_3$ backgrounds, focusing in particular on the low-energy spectrum and its deformations.      
Moreover, once the world-sheet theory is under control, it is possible to perform a systematic analysis of the classical moduli space and to understand whether one-loop tachyons appear, without relying on the approximation to flat space. The latter analysis could be performed by looking at the $2$pt functions of the relevant moduli on the torus. This is not part of the present paper, but we plan to come back to this issue in the near future. 

In addition, the behaviour of non-supersymmetric strings on such AdS$_3$ backgrounds, aside from the bosonic string, is largely unexplored. Hence, we take this opportunity to describe other non-supersymmetric backgrounds. We provide as an example the type 0B superstring, even though the same analysis can be performed for the other non-supersymmetric tachyonic closed string theories. In flat space, it was shown that theories admitting only space-time bosons cannot avoid tachyons from being present in the spectrum \cite{Angelantonj:2023egh,Leone:2023qfd} (see \cite{Dienes:1994np,Cribiori:2020sct,Cribiori:2021txm,Angelantonj:2023egh,Leone:2023qfd} for a more general discussion on the characterisation of tachyons from a world-sheet perspective) but the result is not guaranteed to hold in the case of other non-compact backgrounds. Indeed, in \cite{Eberhardt:2025sbi}, considering backgrounds which correspond to an {\em euclidean} continuation of AdS$_3$, the bosonic string is shown to be tachyon-free. Nonetheless, we verify that, as in flat space, the type 0B superstring on the standard AdS$_3$ backgrounds is tachyonic.  

The paper is organised as follows: in section \ref{sec:sl2Rgeneralities} we shortly review the properties of the (universal cover of the) $\text{SL}(2,\mathbb{R})$ WZW model which describes strings on AdS$_3$. In section \ref{sec:type0} we describe the type 0B superstring quantised on the AdS$_3 \times S^3 \times T^4$  and the  AdS$_3 \times S^3 \times S^3 \times S^1$ backgrounds and we show how tachyons emerge in these situations. Afterwards, in section \ref{sec:16x16}, we describe on the same backgrounds the $\text{Spin}(16) \times \text{Spin}(16) \rtimes \mathbb{Z}_2$ heterotic string and we turn on a specific Wilson line that introduces tachyons in the spectrum. Finally, we provide in section \ref{sec:outlook} future directions that can be explored from this analysis and in Appendix \ref{app:thetaandcharacters} we collect useful properties of the modular functions and characters involved. As an addendum, we describe the partition function of the supersymmetric $\text{E}_8 \times \text{E}_8 \rtimes \mathbb{Z}_2$ heterotic string in Appendix \ref{app:SUSYE8}. 

\section{The \texorpdfstring{$\mathfrak{sl}(\mathfrak{2},\mathbb{R)}$}{sl(2,R)} WZW model}\label{sec:sl2Rgeneralities}

The AdS$_3$ space-time is known to be a group manifold given by the universal cover of $\text{SL}(2,\mathbb{R})$, denoted by $\widetilde{{\text{SL}}(2,\mathbb{R}})$. Hence, the propagation of the bosonic string is described by the associated Wess-Zumino-Witten (WZW) model \cite{Maldacena:2000hw,Maldacena:2000kv,Maldacena:2001km}, which realises a Ka$\check{\text{c}}$-Moody algebra $\widehat{\mathfrak{sl}(\mathfrak{2},\mathbb{R})}_k$ both on the left and right moving sectors given by   
\begin{equation}\label{eq:sl2RKMalgebra}
    \begin{aligned}
       & \big [ J^3_n \, , \, J^3_m \big ]= -\frac{k}{2} n \delta_{n+m} \, ,
        \\
       & \big [ J^3_n \, , \, J^{\pm}_m \big ]= \pm J^{\pm}_{n+m} \, ,
       \\
       & \big [ J^+_n \, , \, J^-_m \big ]= k n \delta_{n+m} - 2 \, J^3_{n+m}\, ,
    \end{aligned}
    \quad
        \begin{aligned}
       & \big [ \bar J^3_n \, , \, \bar J^3_m \big ]= -\frac{k}{2} n \delta_{n+m} \, ,
        \\
       & \big [ \bar J^3_n \, , \, \bar J^{\pm}_m \big ]= \pm \bar J^{\pm}_{n+m} \, ,
       \\
       & \big [ \bar  J^+_n \, , \,\bar  J^-_m \big ]= k n \delta_{n+m} - 2 \, \bar J^3_{n+m}\, ,
    \end{aligned}
\end{equation}
with $\big [ J^a_n \, , \, \bar J^b_m \big ]=0$. The central extension $k$ appearing in the algebra as the $\mathfrak{sl}(\mathfrak{2},\mathbb{R})$ anomaly enters the non-linear sigma model as the flux of the Kalb-Ramond field and corresponds to the level of the affine algebra. The WZW models are known to realise a conformal field theory via the Sugawara construction according to which the energy-momentum tensor reads
\begin{equation}\label{eq:energymomentumtensor}
    T(z)=\frac{1}{2(k+h^\vee)} \sum_{a,b} \kappa_{a,b} : J^a J^b :(z) \, ,
\end{equation}
with $\kappa$ being the Killing form of the underlying finite-dimensional algebra. We are interested in the split form of the $A_1$ algebra, hence the Killing form in the basis $\{ J^+, J^-,J^3 \}$ described in \eqref{eq:sl2RKMalgebra} reads
\begin{equation}
    \kappa= \begin{pmatrix}
        0 & 1 & 0
        \\
        1 & 0 & 0
        \\
        0 & 0 & -2
    \end{pmatrix} \, .
\end{equation}
Therefore, the Virasoro modes become 
\begin{equation}
    L_n= \frac{1}{2(k+h^\vee)} \sum_{m \in \mathbb{Z}} \Big (: J^+_{n-m} J^-_{m} :+: J^-_{n-m} J^+_{m}: - 2: J^3_{n-m} J^3_{m}: \Big) \, ,
\end{equation}
where the normal ordering prescription is needed for the Hamiltonian $L_0-c/24$\footnote{We are implicitly describing the world-sheet as the complex plane.}, the central charge $c= 3 k/(k+h^\vee)$. Given the expression of the Virasoro modes, we can complete the algebra as follows
\begin{equation}
    \begin{aligned}
       & \big [ L_n \, , \, L_m \big ]= (n-m) L_{n+m} + \frac{c}{12} n(n^2-1)\delta_{n+m} \, ,  
       \\
       & \big [ L_n \, , \, J^a_m \big ]=  -n J^a_{n+m}   \, .
    \end{aligned}
\end{equation}
Moreover, the algebra admits an inner automorphism $\sigma_w$ known as the {\em spectral flow}\footnote{One can see the spectral flow in two equivalent ways \cite{Pakman:2003kh}: either as flowed operators acting on the unflowed Hilbert space or as unflowed opertors acting on the flowed vector space.}   \cite{Henningson:1991jc}
\begin{equation}\label{eq:spectralflowbos}
    L_n \to L_n - w J_n^3-  \frac{k \, w^2}{4} \delta_n \, , \ \ J_n^{3} \to J_n^3 + \frac{w \, k}{2} \delta_n \, , \ \ \text{and} \ \ J_n^{\pm} \to J_{n \mp w}^{\pm} \, , 
\end{equation}
with a similar action on the right-moving generators. The global group structure affects the nature of the spectral flow parameter to be imposed on the left and right moving sectors: since the time coordinate is non-periodic $w=\bar w$\footnote{If we are not interested in the universal cover of $\text{SL}(2,\mathbb{R})$ then we simply have to ask that $w-\bar w \in 2 \mathbb{Z}$.} \cite{Maldacena:2000hw}. The spectral flow is present even for compact WZW models, whose action, as will become clear in the following, induces a reshuffling of the original representations. In the non-compact case, however, new kinds of representations emerge \cite{Giribet:2007wp}. The difference in the two settings relies on the nature of the unitary representations: for compact spaces unitary representations are finite-dimensional inducing a reshuffling of the original representations, while for the non-compact case they are infinite-dimensional hence inducing new representations.

The representations of the string spectrum are organised in terms of representations of the affine algebra. These are built via the action of oscillators on the ground states, forming a unitary representation of the zero mode algebra $\mathfrak{sl}(\mathfrak{2},\mathbb{R})$. These representations are well-known. They are labelled by $|j,m\rangle$ related to the eigenvalues of the quadratic Casimir $Q$ and the Cartan generator $J^3_0$
\begin{equation}
    Q|j,m\rangle=j(1-j)|j,m\rangle \, , \qquad J^3_0 |j,m\rangle=m |j,m\rangle \, ,
\end{equation}
where the generator $J^{\pm}_0$ increases(lowers) the Cartan eigenvalue.
These representations are classified into \cite{Kitaev:2017hnr}
\begin{itemize}
    \item {\em the trivial representation}, $\mathbf{1}$: this is a $1d$ representation $|0,0\rangle$ annihilated by both $J^{\pm}_0$. 
    \item {\em the highest weight discrete series}, $D^-_j$: this is a infinite-dimensional representation bounded from above, whose highest weight state satisfies 
    \begin{equation}
        J^+_0 |j,-j\rangle=0 \, .
    \end{equation}
    The possible values of the Cartan eigenvalue $m=-j-\ell$ with $\ell \in \mathbb{N}$ and $0 < j$. 
    \item {\em the lowest weight discrete series}, $D^+_j$: this is an infinite-dimensional representation bounded from below, whose highest weight state satisfies 
    \begin{equation}
        J^-_0 |j,j\rangle=0 \, .
    \end{equation}
    The possible values of the Cartan eigenvalue $m=j+\ell$ with $\ell \in \mathbb{N}$ and $0 < j$. 
    \item {\em the principal continuous series}, $C^\alpha_p$: this is a infinite-dimensional representation which is unbounded with $j=\frac12+ip$, with $p \in \mathbb{R}$, and $m=\alpha+\ell$, with $\alpha \in [0,1)$ and $\ell \in \mathbb{Z}$. In such a case the eigenvalue of the Cartan generator and the Casimir are not related 
    \begin{equation}
        Q|j,m\rangle =\big (\tfrac14+p^2 \big ) |j,m\rangle \, , \qquad J^3_0 |j,m\rangle=(\alpha+\ell) |j,m\rangle \, .
    \end{equation}
    \item  {\em the complementary continuous series}, $\mathcal{E}^\alpha_j$: this is an unbounded representation where $j \in \mathbb{R}$ with $\frac12 < j < 1$ and the Cartan generator eigenvalue is $m=\alpha+\ell$, $\ell \in \mathbb{Z}$ and $j-\frac12<|\alpha-\frac12|$.
\end{itemize}
Despite being unitary, not all representations admit a physical interpretation: only the discrete representations with $\frac12<j$ are square integrable and only the principal series can be normalised to a delta-function (see \cite{Kitaev:2017hnr} for a detailed discussion). Such representations should be tensored with themselves to obtain the correct $\mathfrak{so(2,2)}$ representations. The representations of $\mathfrak{so(2,2)}$ can be obtained by defining the eigenvalues $E,s$ associated with Cartan generators of the algebra. In terms of the decomposition into $\mathfrak{sl}(\mathfrak{2},\mathbb{R}) \oplus \mathfrak{sl}(\mathfrak{2},\mathbb{R})$ algebra we can identify the two labels $E,s$ as
\begin{equation}\label{eq:so22irreps}
    E=m+\bar m \, , \qquad  s=|m-\bar m| \, .
\end{equation}
where $(m,\bar m)$ correspond to the eigenvalues of the Cartan generators of the two $\mathfrak{sl}(\mathfrak{2},\mathbb{R})$  algebras\footnote{These correspond to the conformal weight of the dual CFT.}. 

As mentioned before, the representations of the full affine algebra follow from the action of the oscillators $J^a_{-n}$ on these ground states satisfying the condition $J^a_n |j,m\rangle=0$, for $n>0$. 
However, it is well known that, even though the ground states representations are unitary, negative norm states appear at higher level on the Verma module, due to the action of the Fourier modes of the currents. One can show that these states can be decoupled if the string spectrum is consistently organised in terms of BRST cohomology classes (or equivalently if the Virasoro constraints are satisfied). In the case at hand, this is achieved if we restrict to $0< j \leq k/2$ \cite{Evans:1998qu}, with an internal CFT $X$ that allows to reach criticality $c_X=26-\frac{3k}{k-2}$. In the old covariant language, the states lying in the BRST cohomology classes are those satisfying 
\begin{equation}
    J^3_m |\psi \rangle=0 \, , \quad n>0 \, , 
\end{equation}
so that the unitary spectrum is provided by the gauging of the $\mathfrak{u(1)}$ Cartan sub-algebra \cite{Lykken:1988ut,Hwang:1990aq,Evans:1998qu,Pakman:2003kh}.

The spectral flow induces new representations that have to be taken into account. These representations are obtained from the previous ones with fixed $j$ and $m$ and are mapped to bounded representations \cite{Giribet:2007wp}, even though the unflowed representations are unbounded. This can be seen from the definition of the spectrally flowed operators in \eqref{eq:spectralflowbos}. A given representation of the zero mode algebra $|j,m \rangle $ is a Virasoro primary and under spectral flow
\begin{equation}
    \tilde{J}^{\pm}_{0} |j,m\rangle=J^{\pm}_{\mp w} |j,m\rangle \, , \qquad \tilde{J}^3_0 |j,m\rangle= \big ( J^3_0 + \tfrac{k}{2}w \big ) |j,m\rangle \, .
\end{equation}
We can immediately see that for $w>0$ the new representations are bounded from below, while for $w<0$ the new representations are bounded from above. Since these are annihilated by $\tilde{J}^\pm_0$ these are now highest or lowest weight representations of the zero mode algebra with Cartan eigenvalue for the highest or lowest weight state provided by $m+w\frac{k}{2}$. In the following, we will denote these new representations with an extra label describing the spectral flow parameter $|j,m,w\rangle$, identifying $D_{j,w}^{\pm}$ and $C_{p,w}^\alpha$. In terms of $\mathfrak{so(2,2)}$ representations, the eigenvalues entering the definition of $E$ and $s$ correspond to those of the flowed Cartan generators.
For a particular choice of the spectral flow parameter $w=-1$, the lowest discrete representations are mapped to highest weight discrete representations \cite{Maldacena:2000hw} and hence considering all the values of the $w$ for the lowest weight representations already takes into account all the discrete spectrum.
Reducing the spectrum to the representations admitting at most a $\delta$-function normalisation requires $1/2 <j$ and allowing the spectral flow to be a symmetry of the theory implies $j< (k-1)/2$, so that the spectrum has to satisfy the so-called {\em Maldacena-Ooguri bound} \cite{Maldacena:2000hw}, $1/2 < j <(k-1)/2$. 

The spectrum will be described by using the one-loop partition function \cite{Maldacena:2000hw,Maldacena:2000kv,Israel:2003ry} expressed in terms of the characters $\chi_{j,w}^{\pm}$ and $\chi_{p,w}^{\alpha}$ \cite{Bakas:1991fs,Maldacena:2000hw,Kato:2000tb}, reported for the sake of completeness in the Appendix \ref{app:thetaandcharacters}. However, as pointed out in the previous discussion, the unitary representations appearing at each level of the affine algebra are infinite dimensional, hence the {\em specialised} characters are formally infinite at each level \cite{Petropoulos:1999nc}. This kind of divergence is also present in flat space, where, however the abelian structure of the algebra allows to factorise the overall volume out of the partition function and absorb it in the normalisation constant of the path integral. Here, such a procedure is not possible\footnote{Such a theory can be described in terms of Wakimoto representation involving a $(\beta,\gamma)_{(1,0)}$ system and a linear dilaton CFT with an additional runaway potential. This latter term can be ignored if we consider the region close to the boundary, hence resulting in a free field theory \cite{Giveon:1998ns,deBoer:1998gyt,Giribet:1999ft}. In this region, the string coupling, which can be interpreted as the VEV of the scalar describing the linear dilaton CFT \cite{Eberhardt:2023lwd}, is small, and therefore this representation is allowed in string perturbation theory.} and one should require a more sophisticated regularisation technique. 
In the following, we introduce the refined characters obtained by turning on "chemical potentials" for the Cartan generator $z=e^{2\pi i y}$, with $y=y_1+i y_2$, and the central element of the Ka$\check{\text{c}}$-Moody algebra, $u=u_1+i u_2$, to define $\chi_{j,w}^{\pm}(u,z,\tau)$ and $\chi_{p,w}^{\alpha}(u,z,\tau)$. This choice allows to read the representations occurring at each level of the affine algebra, still preserving the modular properties \cite{Kac:1984mq,Kato:2000tb} as summarised in the Appendix \ref{app:thetaandcharacters}. 
The partition function thus follows from the diagonal combination of the characters \cite{Maldacena:2000hw, Kato:2000tb}, where modular invariance follows from the transformation properties of the elliptic functions reported in Appendix \ref{app:thetaandcharacters}. Hence, the one-loop partition function reads
\begin{equation}\label{eq:bosAdS3}
\begin{aligned}
    \mathcal{Z}&= \int_{\mathcal{F}} d \mu \sum_{w=-\infty}^{\infty} \left \{  \int_{\frac12}^{\frac{k-1}{2}} d j \  \chi_{j,w}^{+}  \, \bar \chi_{j,w}^{+} + \int_{0}^{1} d \alpha \int_{-\infty}^{\infty} d p \ \chi_{1/2+i p,w}^{\alpha}  \, \bar \chi_{1/2+i p,w}^{\alpha} \right \} \mathcal{Z}_{X} \mathcal{Z}_{\text{ghost}} \, ,
    \\
    &=\int_{\mathcal{F}} d \mu  \frac{e^{- \pi k \frac{y_2^2}{\tau_2}-4 \pi k u_2}}{\sqrt{\tau_2}} \left \{  \frac{e^{2 \pi \frac{y_2^2}{\tau_2}} }{\big | \Jtheta{1/2}{1/2}{y}{\tau} \big |^2}  + \frac{\sum_{w, \ell} \delta^{(2)}(y+w \tau+\ell) }{\big | \eta(\tau ) \big |^6}  \right \} \mathcal{Z}_{X} \mathcal{Z}_{\text{ghost}} \, .
\end{aligned}
\end{equation}
Here, we have omitted the chemical potentials $y,u$ entering the definition of the characters to lighten the notation.
We have denoted $d\mu = d^2 \tau/\tau_2^2$ the modular invariant measure and $\mathcal{F}$ the fundamental domain. Now that the zero modes are integrated out, we can read the spectrum following the same steps of the analysis on flat space: we impose the mass shell condition from which we can read the allowed values of $j,w$ and looking at the powers of $z$ we can read the representations occurring at that level. This will be explicitly done in the following.

\subsection{The notion of mass} \label{ssec:mass}

In order to discuss the presence of tachyons and the properties of the string spectrum we need to discuss the notion of mass for AdS$_3$ backgrounds. Indeed, this is known to be ambiguous for curved space-times since its definition may vary according to the coupling of the field of interest with the AdS curvature. Nonetheless, following \cite{Ferrara:1998jm,Ferrara:1998ej}, the notion of a gauge field can be defined unambiguously and hence we can define the mass in such a way that gauge fields are massless. This means that the mass (in units of curvature radius) reads \cite{Aharony:1999ti}
\begin{equation} \label{eq:mass}
    m^2=-Q_{\mathfrak{so(2,2)}}- 2s (s-1) \, ,
\end{equation}
where $Q_{\mathfrak{so(2,2)}}$ is the Casimir of the $\mathfrak{so(2,2)}$ algebra. For highest or lowest weight representations, the Casimir reads \cite{Aharony:1999ti}
\begin{equation}
    Q_{\mathfrak{so(2,2)}}= -E(E-2)-s^2 \, ,
\end{equation}
where $E$ and $s$ correspond to the Cartan eigenvalues of the associated extremal weights defining the representation. For all these representations, one can show that the Breitenlohner-Freedman (BF) bound
\begin{equation}\label{eq:BFbound}
    m^2 \geq -(s-1)^2 \, ,
\end{equation}
can never be violated as $E$ and $s$ are real quantities.
On the other hand, the situation for the unflowed continuous representations is more subtle since these representations are unbounded. These representations are puzzling in the dual holographic theory and are associated with the presence of tachyons. Indeed, from a pure algebraic perspective, inserting into \eqref{eq:mass} the decomposition of the Casimir in terms of those of the two $\mathfrak{sl(2,}\mathbb{R} \mathfrak{)}$ for the unflowed continuous representations implies that the bound \eqref{eq:BFbound} is always violated\footnote{Recall that we are considering irreducible representations $p \neq 0$ since at $p=0$ it decomposes into discrete representations \cite{Maldacena:2000hw}.}. 

\section{The type 0 superstring}\label{sec:type0}

The Ramond-Neveu-Schwarz superstring is obtained as the supersymmetric extension of the bosonic string theory on the 2d world-sheet. In the superconformal gauge, this means introducing additional world-sheet fermions realising the $\mathcal{N}=(1,1)$ supersymmetry on the world-sheet. It is well known that one-loop modular invariance, and hence the locality of the world-sheet theory, requires a proper choice of the {\em GSO projection} \cite{Gliozzi:1976jf,Gliozzi:1976qd}. The choice of the projection corresponds to different ways to sum over all the spin structures of the world-sheet fermions \cite{Seiberg:1986by} admitting four consistent possibilities: two lead to a space-time supersymmetric theory, the type IIA and the type IIB superstring, while the remaining ones to non-supersymmetric theories, the type 0A and the type 0B superstring (see \cite{Angelantonj:2002ct,Leone:2025mwo,Dudas:2025ubq} for reviews). In this Section, we will focus on the latter.

For AdS$_3$ backgrounds, world-sheet fermions introduce additional subtleties. In particular, fermions transform in the adjoint representation of the $\mathfrak{sl}(\mathfrak{2},\mathbb{R})$ algebra, implying that
\begin{equation}
   \begin{matrix} 
   \begin{aligned}
        &\big [ J^3_n \, , \, \psi^3_r \big ]= 0 \, ,
        \\
        &\big [ J^3_n \, , \, \psi^{\pm}_r \big ]= \pm \psi^{\pm}_{n+r} \, , 
       \\
        &\big [ J^{\pm}_n \, , \, \psi^{\mp}_r \big ]= \mp 2 \psi^3_{n+r} \, ,
    \end{aligned}
    &
       \begin{aligned}
        &\big \{ \psi^3_r \, , \, \psi^3_s \big \}= -\frac{k}{2}  \delta_{r+s} \, ,
        \\
        & \big \{ \psi^{+}_r \, , \, \psi^{-}_s \big \}= k \delta_{r+s} \, ,
       \\
       & \big [  J^{\pm}_n \, , \, \psi^3_r \big ]=\mp  \psi^{\pm}_{n+r} \, .
    \end{aligned}
    \end{matrix}
\end{equation}
Because of the presence of world-sheet fermions, the $\widehat{\mathfrak{sl}(\mathfrak{2},\mathbb{R})}_k$ Ka$\check{\text{c}}$-Moody algebra is enhanced to a superalgebra $\widehat{\mathfrak{sl}(\mathfrak{2},\mathbb{R})}_k^{(1)}$.
However, in contrast to the flat space case, fermions are coupled to Ka$\check{\text{c}}$-Moody currents, forbidding in this way the factorisation of the Hilbert space between bosonic and fermionic oscillators. Nonetheless, it is possible to redefine the currents in such a way to decouple the world-sheet fermions \cite{Giveon:1998ns} 
\begin{equation}
    \begin{aligned} \label{eq:WZWshift}
        &\mathcal{J}^{\pm}(z)= J^{\pm}(z) \pm \frac{2}{k} : \psi^3 \psi^{\pm}:(z) \, ,
        \\
        &\mathcal{J}^{3}(z)= J^{3}(z) + \frac{1}{k} : \psi^+ \psi^-:(z) \, .
    \end{aligned}
\end{equation}
at the price of shifting the level of the algebra by $k \to k -h^\vee$, hence $k \to k+2$ for $\mathfrak{sl}(\mathfrak{2},\mathbb{R})$. Indeed, the super-affine Ka$\check{\text{c}}$-Moody algebra becomes
\begin{equation}
\begin{matrix}
\begin{aligned}
    &\big [ \mathcal{J}^3_n, \mathcal{J}^3_m \big ] = -\frac{k+2}{2} n \delta_{n+m} \, , 
    \\
    & \big [ \mathcal{J}^3_n, \mathcal{J}^{\pm}_m \big ] = \pm \mathcal{J}^{\pm}_{n+m} \, , 
    \\
     &\big [ \mathcal{J}^+_n, \mathcal{J}^{-}_m \big ] =   (k+2) n \delta_{n+m} - 2 \mathcal{J}^3_{n+m} \, , 
     \end{aligned}
     &
     \begin{aligned}
     & \big [ \mathcal{J}^a_n, \psi^b_r \big ] = 0 \, ,
    \\
     & \big \{ \psi^3_r \, , \, \psi^3_s \big \}= -\frac{k}{2}  \delta_{r+s} \, ,
    \\
      & \big \{ \psi^{+}_r \, , \, \psi^{-}_s \big \}= k \delta_{r+s} \, .
     \end{aligned}
\end{matrix} 
\end{equation} 
The energy-momentum tensor and the supercurrent then become
\begin{equation}
    T(z) \to \frac{1}{2k} \sum_{a,b} \kappa_{a,b} \Big ( : \mathcal{J}^a \mathcal{J}^b:(z)+ : \psi^a \partial \psi^b:(z) \Big )\, ,
\end{equation}
and
\begin{equation}\label{eq:supercurrent}
    G(z)= \frac{1}{k} \Big ( \sum_{a,b} \kappa_{a,b}  :\mathcal{J}^a \psi^b:(z)- \frac{i}{6 k} \sum_{a,b,c} f_{abc} :\psi^a \psi^b \psi^c: (z) \Big ) \, ,
\end{equation}
following from the general expression of the shift 
\begin{equation}
    \mathcal{J}^a=J^a +\frac{i}{2k} \sum_{b,c} f^a_{bc} :\psi^b \psi^c:(z) \, . 
\end{equation}
In what follows, we will be interested in backgrounds of the type AdS$_3 \times S^3 \times X$, with $X= T^4$ or $X= S^3 \times S^1$ such that they saturate the central charge. The torus contribution is made of free bosons and fermions, while the $S^3$ sigma model is described by a superaffine algebra given by the product of two copies of $\widehat{\mathfrak{su(2)}}^{(1)}_{k}$.
For the latter, as in the $\widehat{\mathfrak{sl}(2,\mathbb{R})}_k^{(1)}$ case, we have an interacting bosonic theory coupled to three world-sheet fermions. A similar story applies for these world-sheet fermions, where the level of the underlying $\widehat{\mathfrak{su(2)}}_k$ algebra is shifted by $k \to k-2$. This then means that the $S^3$ sigma-model contributes with three free fermions and form finite-dimensional representations of $\mathfrak{su(2)}$. 
In the one-loop partition function, the $S^3$ sigma model then contributes with the characters associated with the $\widehat{\mathfrak{su(2)}}_{k-2}$ along with the $\widehat{\mathfrak{so(3)}}_1$ characters describing the world-sheet fermions. In the following, we will consider the diagonal modular invariant since we are keeping $k$ unspecified, and this is the only modular invariant combination existing for any value of $k$. Moreover, the sum over the spin structures giving rise to the $\widehat{\mathfrak{so(3)}}_1$ algebra is taken in such a way to reflect the choice of the GSO projection characterising the type 0B superstring.  

For the $\mathfrak{sl}(\mathfrak{2},\mathbb{R})$ case, we also need to discuss the effect of the spectral flow on the world-sheet fermions. 
Since we have decoupled the  $\mathfrak{sl}(\mathfrak{2},\mathbb{R})$ currents from the world-sheet fermions, preserving $N=1$ world-sheet supersymmetry requires performing a spectral flow on the fermions as well.  
Flowing the world-sheet fermions however implies flowing the world-sheet fermion number $F \to F+w$, along with the generators \cite{Giribet:2007wp}. This means that, following \cite{Maldacena:2000hw}, the contribution  of world-sheet fermions with given spin structures, with the Cartan direction eliminated by the superghosts, reads
\begin{equation}
    \begin{aligned}
    \mathcal{Z}^w_{\psi} \big [ \substack{\alpha \\ \beta} \big ] &= \text{tr}_{\alpha} \ e^{2\pi i\beta (F+ w)} q^{L_0-w J^3_0 -\frac{k w^2}{4}-\frac{c}{24}} z^{J^3_0+ \frac{k w}{2}} \, 
        \\
        &= i^{\frac32} q^{-\frac{(k+2)w^2}{4}} z^{\frac{(k+2)w}{2}} \frac{\Jtheta{\alpha}{\beta}{y}{\tau}}{\eta} \, ,
\end{aligned}
\end{equation}
as more explictly shown in Appendix \ref{app:thetaandcharacters}.
The spectral flow enters in the partition by contributing to the zero point energy but it does not appear in the contribution coming from the oscillators\footnote{If one wants to extract the degrees of freedom identifying the dual CFT it is convenient to reorganise the spectrum in a different way, where the degrees of freedom contributing to the space-time partition function are manifest following \cite{Gaberdiel:2018rqv}.}. Hence, we can write down the partition function of superstring theories using expressions in terms of the $\widehat{\mathfrak{so(8)}_1}$ characters suitably decomposed. This is indeed consistent with the result discussed in \cite{Israel:2003ry}, where the contribution from the trace over the world-sheet fermions is organised in terms of theta functions with no spectral flow \footnote{The effect is similar to what happens for the world-sheet bosons, where the oscillator contribution can be reorganised as $\Jtheta{1/2}{1/2}{y}{\tau}$.}.

We are now in the position to compute the spectrum of the type 0B superstring from the expressions of the partition function for the AdS$_3 \times S^3 \times T^4$ and AdS$_3 \times S^3 \times S^3 \times S^1$ for generic levels of the underlying WZW models. We proceed by analysing the two cases separately. 

\subsection{AdS\texorpdfstring{$_3 \times S^3 \times T^4$}{3 x S3 x T4}}

For such a background, the cancellation of the conformal anomaly requires
\begin{equation}\label{eq:criticalitytype0T4}
    \frac{3 (k+2)}{k} +  \frac{3 (k'-2)}{k'} + \frac32+\frac32 + 4 + \frac42=15 \, ,
\end{equation}
implying $k=k'$\footnote{In the tensionless limit for AdS$_3 \times S^3 \times T^4$ background, occurring for $k=1$, the level becomes negative introducing additional difficulties. In this context it is more suited to use the hybrid formalism \cite{Berkovits:1999im} which is manifestly space-time supersymmetric. However, in this paper, space-time supersymmetry is absent and we are not interested in this limit. Hence, we will stick to the RNS superstring.}. The partition function for the type 0B superstring is fully specified once the following GSO projection is provided
\begin{equation}\label{eq:0BGSOprojection}
    \bigoplus_{\alpha=\text{NS},\text{R}} \text{tr}_{\alpha, \bar \alpha} \bigg (\frac{1+(-1)^F}{2} \bigg ) \bigg (\frac{1 +(-1)^{\bar F}}{2} \bigg ) \oplus \text{tr}_{\alpha, \bar \alpha} \bigg (\frac{1-(-1)^F}{2} \bigg ) \bigg (\frac{1-(-1)^{\bar F}}{2}\bigg )
\end{equation}
with $F(\bar F)$ denoting the world-sheet fermion number acting on the (anti-)holomorphic sector. 
Therefore we can write down the partition function 
\begin{equation}\label{eq:toruspartitionfunctionT^4}
\begin{aligned}
     \mathcal{T}_{0\text{B}}(T^4)&= \int_{\mathcal{F}} d \mu \frac{e^{- \pi k \frac{y_2^2}{\tau_2}-4 \pi k u_2}}{\sqrt{\tau_2}} \bigg \{  \frac{e^{2 \pi \frac{y_2^2}{\tau_2}} }{\big | \Jacobitheta{1/2}{1/2} \big |^2}  + \frac{\sum_{w ,\ell}\delta_{w,\ell} }{\big | \eta \big |^6} \bigg \}\sum_{\ell=0}^{\frac{k-2}{2}} |\chi_{\ell}|^2 \ \Gamma_{(4,4)} \big ( \eta \bar \eta \big )^2
     \\
     & \qquad \qquad \times \, \frac{e^{-8 \pi u_2}}{2}\sum_{a,b=0,\frac12} \frac{\Jacobitheta{a}{b}^4}{ \eta^4}\frac{\Jacobibartheta{a}{b}^4}{ \bar \eta^4} 
     \\
     &= \int_{\mathcal{F}} d \mu \, B(\tau,\bar \tau \, | \, T^4) F_{0\text{B}}(\tau,\bar \tau \,| \,T^4) \, ,
\end{aligned}
\end{equation}
where we have used to lighten the notation $\delta_{w,\ell}(y|\tau)=\delta^{(2)}(y+w \tau+\ell)$ and we have identified the first line of the first expression with the contribution coming from the world-sheet bosons and the $(b,c)_{(2,-1)}$ system, $B(\tau,\bar \tau |T^4)$. In the latter, the term $\Gamma_{(4,4)}$ is the lattice term and may in general depend on the metric and the Kalb-Ramond of the internal torus $\Gamma_{(4,4)}(G,B)$ and we have used the definition of the $\widehat{\mathfrak{su(2)}}_k$ characters reported in the Appendix \ref{app:thetaandcharacters}. Notice that using the modular properties of the functions involved this piece is modular invariant by itself. 
The second line of the first expression has been identified with the contribution coming from the world-sheet fermions. In the first theta-functions we have omitted the dependence on the chemical potential $z=e^{2\pi i y}$ to lighten the notation. Moreover, we have already included the superghosts which compensate the contributions corresponding to the Cartan directions in $\text{SL}(2,\mathbb{R})$ and $\text{SU}(2)$. For our purpose it is instructive to express their contributions in terms of characters as 
\begin{equation}
\begin{aligned}
    F_{0\text{B}}(\tau, \bar \tau \, | \, T^4)=  \, e^{-8 \pi u_2} & \Big \{   \big |  O_2 \ O_2 \ O_4 + O_2 \ V_2 \ V_4+ V_2 \ V_2 \ O_4+ V_2 \ O_2 \ V_4 \big |^2
        \\
        & + \big | V_2 \ O_2 \ O_4 + V_2 \ V_2 \ V_4+ O_2 \ V_2 \ O_4+ O_2 \ O_2 \ V_4 \big |^2
        \\
        & + \big | S_2 \ S_2 \ S_4 + S_2 \ C_2 \ C_4+ C_2 \ S_2 \ C_4+ C_2 \ C_2 \ S_4 \big |^2
        \\
        & + \big | C_2 \ S_2\  S_4 + C_2 \ C_2 \ C_4+ S_2 \ S_2 \ C_4+ S_2 \ C_2 S_4  \big |^2 \Big \} \, ,
    \end{aligned}
\end{equation}
where the first $\widehat{\mathfrak{so(2)}}_1$ characters, whose definition is reported in Appendix \ref{app:thetaandcharacters}, refer to the fermions of AdS$_3$, and thus implicitly carry a dependence on $z=e^{2 \pi i y}$, the second ones to those of $S^3$ and the $\widehat{\mathfrak{so(4)}}_1$ characters to those of $T^4$. The factor in front is inherited by the chemical potential associated with the central extension of the $\widehat{\mathfrak{sl(2,\mathbb{R})}}_{k+2}$ algebra\footnote{Notice that in the bosonic theory this is automatically absent since we have no level shift due to the fermions.} and it is essential to guarantee the modular invariance of this sector since now the characters associated with the AdS$_3$ fermions have a non-trivial dependence on the chemical potential and hence their modular properties are altered according to the formulas in \ref{app:thetaandcharacters}.

Once the partition function is known, we can read from the $q-$expansion the level of the oscillators and we can solve the mass-shell condition in terms of the spin $j$ of $\mathfrak{sl}(\mathfrak{2},\mathbb{R})$. To keep track of the representations of $\mathfrak{su(2)}$, we can turn on a chemical potential for these characters\footnote{This automatically means that to preserve the modular properties of the expressions we need to turn on a chemical potential $e^{2 \pi i u}$ in a similar fashion as the analysis for $\mathfrak{sl(2,\mathbb{R})}$.} too corresponding to $z'=e^{2 \pi i y'}$. Then, by looking at the powers of $z$ and $z'$ we can read the representations that appear at that level, and finally the complete spectrum is obtained by imposing level-matching.    
In general, the mass-shell condition reads
\begin{equation} \label{eq:massshellT4}
    \big (\sigma_w(L_0) - \nu\big ) |\psi\rangle= \bigg (  \frac{\mathcal{Q}}{k} + \frac{\mathcal{Q}_{\mathfrak{su(2)}}}{k} + h_{T^4} +N - w J_0^3-  \frac{k \, w^2}{4} -\nu \bigg ) | \psi \rangle= 0 \, , 
\end{equation}
with  $\nu$ denoting the zero point energy of the NS ($\nu =\frac12$) and R ($\nu=0$) ground states. There are four situations that we have to analyse provided by flowed and unflowed discrete representations and unflowed or flowed continuous representations.

Even though in principle the partition function allows access to all states, in the following we will be interested in the low-energy spectrum\footnote{Note that here the representations occurring at higher levels are still described in terms of $\mathfrak{sl}(\mathfrak{2},\mathbb{R})$ representations. This is different from the flat space scenario where the little group for massive representations is different from the massless, one hence requiring a more sophisticated technology to being able to express the spectrum content in an efficient and covariant fashion \cite{Markou:2023ffh,Basile:2024uxn,Markou:2025xpf}.}.
This is extracted by looking at the lowest level in the underlying world-sheet CFT modules with KK momenta and winding associated with the internal $T^4$ set to 0. Indeed, the $N=0$ level in the NS sector identifies the presence of tachyons in the string spectrum, while for the $N=\frac12$ in the NS sector and the $N=0$ level in the R sector the $\mathfrak{sl}(\mathfrak{2},\mathbb{R})$ spin does not depend on the curvature of AdS, $R \sim \alpha' \sqrt{k}$ \cite{Kutasov:1999xu}.  

Starting form the discrete unflowed representations, taking the $N=\frac12$ level with $\nu=\frac12$ and $N=0$ with $\nu=0$, the mass-shell condition becomes
\begin{equation} \label{eq:conditionunflowedT4}
    \frac{j(1-j)}{k} + \frac{j'(j'+1)}{k} =0 \, ,
\end{equation}
which admits as a solution $j=j'+1$ \cite{Ferreira:2017pgt}, which does not depend on the AdS curvature. 
Moreover, the particle content of the $N=\frac12$ states can be read from the following expansion of characters
\begin{equation}
    \begin{aligned} \label{eq:V8expansionT4}
        &V_2 (O_2 O_4+V_2 V_4) \sim \big ( (z^{-1}+z) q^{\frac12} +\ldots \big )\Big [ 1+ (1+6)q +\ldots +(z'^{-1}+z') 4 q +\ldots \Big ] \, ,
        \\
        & O_2 (V_2 O_4+O_2 V_4) \sim \big ( 1+q +\ldots \big ) \Big [ (z'^{-1}+z') q^{\frac12} +\ldots + 4 q^{\frac12} +\ldots \Big ] \, ,
    \end{aligned}
\end{equation}
which combined with the $|j;j-1;0,0\rangle_{0,0}$ ground state identifies the states
\begin{equation}
\begin{aligned}\label{eq:NSlevel12T4}
    &|j-1;j-1;0,0\rangle_{0,0} \oplus |j+1;j-1;0,0\rangle_{0,0} \oplus |j;j;0,0\rangle_{0,0} 
    \\
    &\qquad \qquad \qquad \oplus |j;j-2;0,0\rangle_{0,0} \oplus |j;j-1;\tfrac12,\tfrac12\rangle_{0,0} \, .
    \end{aligned}
\end{equation}
In the notation above, we have denoted the $\mathfrak{sl}(\mathfrak{2},\mathbb{R})$ spin $j$, the $\mathfrak{su(2)}$ spin $j'$ and the $\mathfrak{so(4)}=\mathfrak{su(2)}\oplus \mathfrak{su(2)}$ representations $(s_1,s_2)$ by $|j,j';s_1,s_2\rangle_{m,n}$. The $N=0$ states in the R sector can be read from
\begin{equation}
    \begin{aligned}\label{eq:S8expansionT4}
        & S_2(S_2S_4 +C_2 C_4) \sim q^{\frac12} (z^{1/2} + \ldots )\Big [ ( z'^{1/2} +\ldots )( 2_s +\ldots  ) + ( z'^{-1/2} +\ldots )( 2_c +\ldots  ) \Big ] \, ,
        \\
        &C_2(S_2 C_4 +C_2 S_4) \sim q^{\frac12} (z^{-1/2} + \ldots )\Big [ ( z'^{1/2} +\ldots )( 2_c +\ldots  ) + ( z'^{-1/2} +\ldots )( 2_s +\ldots  ) \Big ] \, ,
    \end{aligned}
\end{equation}
giving 
\begin{equation}
\begin{aligned}
    &|j+\tfrac12;j-\tfrac12;\tfrac12,0\rangle_{0,0} \oplus  |j+\tfrac12;j-\tfrac32;0,\tfrac12\rangle_{0,0}
    \\
    &\oplus  |j-\tfrac12;j-\tfrac12;0,\tfrac12\rangle_{0,0} \oplus |j-\tfrac12;j-\tfrac32;\tfrac12,0\rangle_{0,0}  \, .
\end{aligned}
\end{equation}
The statements above hold for $j \neq 1$, since for $j=1$ in the NS sector \eqref{eq:NSlevel12T4}, the infinite dimensional representation with $j=0$ is reducible \cite{Ferreira:2017pgt}
\begin{equation}\label{eq:j0representationsdecomposition}
    D_0^+= \mathbf{1} \oplus D_1^+ \, . 
\end{equation}
Indeed, one can show that the $D_0^+$ representation occurring at level $1$ in the $\hat{D}_1^+$ affine module has a null state obtained applying $\mathcal{J}^+_0$ to $\psi^-_{-1/2}|1;0\rangle$ \cite{Ferreira:2017pgt}. 
This extra state provide what we need to obtain the representation with $\mathfrak{su(2)}$ spin $j'=1$, so that the spectrum for $j=1$ reads in the NS sector
\begin{equation}
    |0;0;0,0\rangle_{0,0} \oplus |2;0;0,0\rangle_{0,0} \oplus |1;1;0,0\rangle_{0,0} \oplus |1;0;\tfrac12,\tfrac12\rangle_{0,0} \, .
\end{equation}
In the R sector, we apparently have only one component of the $\mathfrak{su}(2)$ representations by 
\begin{equation}\label{eq:j1Rsector}
    |\tfrac12;\tfrac12,\tfrac12;\tfrac12,0\rangle_{0,0} \oplus  |\tfrac12;\tfrac12,-\tfrac12;0,\tfrac12\rangle_{0,0} \oplus  |\tfrac32;\tfrac12,\tfrac12;\tfrac12,0\rangle_{0,0} \oplus  |\tfrac32;\tfrac12,-\tfrac12;0,\tfrac12\rangle_{0,0} \, ,
\end{equation}
where for the second $\mathfrak{su(2)}$ algebra we have specified also the Cartan eigenvalue. The two components with the same $j=\frac12,\frac32$ hence form the two components of a doublet of $\mathfrak{su(2)}$ as they are conjugate to each other. Indeed, this also follows from the decomposition of $\widehat{\mathfrak{su(2)}}_2 =\widehat{\mathfrak{so(3)}}_1$ characters in terms of parafermions \cite{Gepner:1986hr}, from which we can factor out the $\mathfrak{u(1)}$ factor to be compensated by the ghosts. 

These are the sectors that are shared with the type IIB superstring and indeed if one chooses the corresponding GSO projection 
\begin{equation}\label{eq:GSOIIB}
     | \, \text{tr}_{\text{NS}} (1-(-1)^F) -     \text{tr}_{\text{R}} (1+(-1)^F) \, |^2 \, ,
\end{equation}
the world-sheet fermions contribution to the partition function becomes 
\begin{equation}
\begin{aligned}
    F_{II\text{B}}(\tau, \bar \tau | T^4)=   \, e^{-8 \pi u_2}  & \big | V_2 \ O_2 \ O_4 + V_2 \ V_2 \ V_4+ O_2 \ V_2 \ O_4+ O_2 \ O_2 \ V_4
        \\
        &  - S_2 \ S_2 \ S_4 - S_2 \ C_2 \ C_4- C_2 \ S_2 \ C_4- C_2 \ C_2 \ S_4 \big |^2 \, ,
    \end{aligned}
\end{equation}
consistently with \cite{Israel:2003ry}. From this expression and the previous discussion, we can organise the spectrum taking into account the Madalcena-Ooguri and unitarity bounds as follows
\begin{equation}
    \bigoplus_{j \in I} \Big ( (j-1)_{\text{short}} \oplus (j)_{\text{short}} \oplus 2(j-\tfrac12)_{\text{short}} \Big ) \otimes \overline{\Big ( (j-1)_{\text{short}} \oplus (j)_{\text{short}} \oplus 2(j-\tfrac12)_{\text{short}} \Big ) } \, ,
\end{equation}
where $I=\{ j  \in \mathbb{Z}/2 \, , \, j\geq 1, \, j \leq \frac{k}{2} \}$ and $(j)_{\text{short}}$ is a short supermultiplet for the small $\mathcal{N}=4$ superalgebra $\mathfrak{psu(1,1|2)}$, namely  in term of $\mathfrak{sl}(\mathfrak{2},\mathbb{R}) \otimes \mathfrak{su(2)}$ representations $(j)_{\text{short}}= |j;j\rangle \oplus 2 |j+\frac12;j-\frac12\rangle \oplus |j+1;j-1\rangle$. For $j=1$, the spectrum is truncated since representations with highest weight $j=0$ and $j=\frac12$ appear and hence we have
\begin{equation}
    (0,0)_{\text{short}} \oplus 2 \cdot (\tfrac12,\tfrac12)_{\text{short}} \oplus (1,1)_{\text{short}} \otimes  \overline{(0,0)_{\text{short}} \oplus 2 \cdot (\tfrac12,\tfrac12)_{\text{short}} \oplus (1,1)_{\text{short}}} \, ,
\end{equation}
where $ (0,0)_{\text{short}} =|0;0\rangle$ and $(\tfrac12,\tfrac12)_{\text{short}}=|\frac12;\frac12 \rangle \oplus 2 |0;0\rangle$. Hence, this is precisely the spectrum described in \cite{Ferreira:2017pgt,Eberhardt:2017pty} which reproduces the supergravity description in \cite{deBoer:1998kjm}.

\begin{figure}
    \centering
    \includegraphics[width=0.5\linewidth]{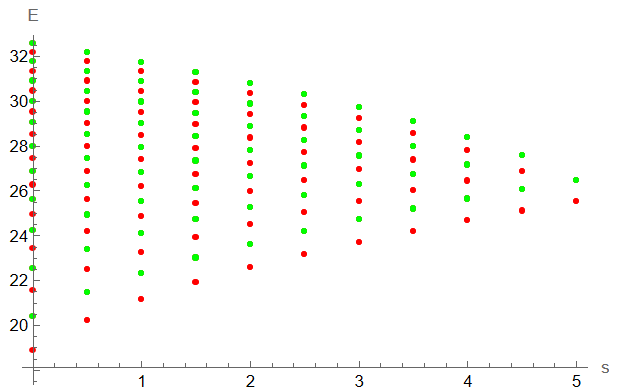}
    \caption{We display the dispersion relation $(E,s)$ for the discrete representations choosing for simplicity $h_{T^4}=\bar h_{T^4}=q=\bar q=j'=0, w=-2,k=7$. The points in green describe the states appearing both in the type IIB and type 0B superstring, while the points in red correspond to the additional sector entering the type 0B superstring.}
    \label{fig:discretespectrum}
\end{figure}

However, for the type 0B superstring we have additional contributions. Indeed, in the NS sector,
\begin{equation}
\begin{aligned} \label{eq:O8expansionT4}
    &O_2(O_2 O_4 +V_2 V_4) \sim (1+q+\ldots) \Big [ (1+q+\ldots)(1+6q+\ldots ) 
    \\
    & \qquad \qquad \qquad \qquad \qquad \qquad \qquad+( ( z'^{-1}+z')q^{\frac12}+\ldots)(4 q^{\frac12} +\ldots ) \Big ] \, ,
    \\
    &V_2 (V_2 O_4 + O_2 V_4 ) \sim ( ( z^{-1}+z)q^{\frac12}+\ldots) \Big [ ( z'^{-1}+z')q^{\frac12}+\ldots)( 1+ 6q +\ldots ) 
    \\
    & \qquad \qquad \qquad \qquad \qquad \qquad \qquad+ (1+q+\ldots)(4 q^{\frac12} +\ldots ) \Big ] \, ,
    \end{aligned}
\end{equation}
from which we see that we have a state at $N=0$ level, while all the other contributions occur at $N\geq 1$. At $N=0$ in the NS sector the mass-shell condition becomes
\begin{equation}
    \frac{j(1-j)}{k}+\frac{j'(1+j')}{k} -\frac12=0 \, ,
\end{equation}
which admits as a solution
\begin{equation}
    j=\frac12+ \sqrt{j'(1+j')+\frac{1-2k}{4}} \, ,
\end{equation}
constrained by $j'\geq \sqrt{k/2}-1/2$. Hence they contribute to the spectrum as
\begin{equation}
    \bigotimes_{\substack{\frac12 <j<\frac{k+1}{2} \\  \frac{\sqrt{2k}-1}{2}\leq j'\leq \frac{k-2}{2}}} |j;j'\rangle \otimes  \overline{{|j;j'\rangle}} \, , \qquad j=\frac12+ \sqrt{j'(1+j')+\frac{1-2k}{4}} \, .
\end{equation}  
However, these states correspond to massive scalars since, being discrete representations, the mass can never violate the BF bound and correspond to massive string states. Nonetheless, such $N=0$ states will be responsible for the appearance of tachyons when we will analyse the contributions from the continuous representations.  
In addition, we have another contribution from the R sector which is given by
\begin{equation}
    \begin{aligned} \label{eq:C8expansionT4}
        & S_2(S_2 C_4 +C_2 S_4)  \sim q^{\frac12} (z^{1/2} + \ldots )\Big [ ( z'^{1/2} +\ldots )( 2_c +\ldots  ) + ( z'^{-1/2} +\ldots )( 2_s +\ldots  ) \Big ] \, ,
        \\
        &C_2 (S_2S_4 +C_2 C_4)\sim q^{\frac12} (z^{-1/2} + \ldots )\Big [ ( z'^{1/2} +\ldots )( 2_s +\ldots  ) + ( z'^{-1/2} +\ldots )( 2_c +\ldots  ) \Big ] \, ,
    \end{aligned}
\end{equation}
hence providing 
\begin{equation}
\begin{aligned}
    &|j+\tfrac12;j-\tfrac12;0,\tfrac12\rangle_{0,0} \oplus  |j+\tfrac12;j-\tfrac32;\tfrac12,0\rangle_{0,0} 
    \\
    &\oplus  |j-\tfrac12;j-\tfrac12;\tfrac12,0\rangle_{0,0} \oplus |j-\tfrac12;j-\tfrac32;0,\tfrac12\rangle_{0,0}  \, .
\end{aligned}
\end{equation}
Still, in the $j=1$ case the contribution is given by
\begin{equation}
    |\tfrac12;\tfrac12,\tfrac12;0,\tfrac12\rangle_{0,0} \oplus  |\tfrac12;\tfrac12,-\tfrac12;\tfrac12,0\rangle_{0,0} \oplus  |\tfrac32;\tfrac12,\tfrac12;0,\tfrac12\rangle_{0,0} \oplus  |\tfrac12;\tfrac12,-\tfrac12;\tfrac12,0\rangle_{0,0} \, .
\end{equation}
Putting everything together we can write down the spectrum as 
\begin{equation}
\begin{aligned}
    &\bigoplus_{j\in I}  \Big ( |j-1;j-1\rangle \oplus |j+1;j-1\rangle \oplus
    |j;j-2\rangle \oplus |j;j\rangle \oplus 4 |j;j-1\rangle \Big ) 
    \\
    & \qquad\otimes \overline{\Big ( |j-1;j-1\rangle \oplus |j+1;j-1\rangle \oplus
    |j;j-2\rangle \oplus |j;j\rangle \oplus 4 |j;j-1\rangle \Big )} \, ,
    \end{aligned}
\end{equation}
from the NS sector and
\begin{equation}
\begin{aligned}
  &\bigoplus_{j \in I} \bigg \{ \Big ( 2 |j+\tfrac12;j-\tfrac12\rangle \oplus 2 |j+\tfrac12;j-\tfrac32\rangle \oplus 2 |j-\tfrac12;j-\tfrac12\rangle \oplus 2|j-\tfrac12;j-\tfrac32\rangle \Big )
  \\
  &\quad \otimes \overline{\Big ( 2 |j+\tfrac12;j-\tfrac12\rangle \oplus 2 |j+\tfrac12;j-\tfrac32\rangle \oplus 2 |j-\tfrac12;j-\tfrac12\rangle \oplus 2|j-\tfrac12;j-\tfrac32\rangle \Big )}
  \\
  & \oplus \Big ( 2 |j+\tfrac12;j-\tfrac12\rangle \oplus 2 |j+\tfrac12;j-\tfrac32\rangle \oplus 2 |j-\tfrac12;j-\tfrac12\rangle \oplus 2|j-\tfrac12;j-\tfrac32\rangle \Big )
  \\
  &\quad \otimes \overline{\Big ( 2 |j+\tfrac12;j-\tfrac12\rangle \oplus 2 |j+\tfrac12;j-\tfrac32\rangle \oplus 2 |j-\tfrac12;j-\tfrac12\rangle \oplus 2|j-\tfrac12;j-\tfrac32\rangle \Big )} \bigg \} \, ,
  \end{aligned}
\end{equation}
from the R sector. 
We can discuss in detail what happens for the case $j=1$. This case is interesting from a physical point of view since it encodes massless fields according to the discussion in \ref{ssec:mass}. Moreover, one can show that no massless states may arise from other representations if $k>1$\footnote{If $k=1$ additional massless fields arise from the spectrally flowed continuous representations denoting the tensionless limit \cite{Gaberdiel:2018rqv}.}. In particular, we have two possible situations in which we can have massless fields: $E=s$ and $E=2$, $s=0$. This means that the states 
\begin{equation}
\begin{aligned}
    &\Big (|0;0;0,0\rangle_{0,0} \oplus |2;0;0,0\rangle_{0,0} \oplus |1;1;0,0\rangle_{0,0} \oplus |1;0;\tfrac12,\tfrac12\rangle_{0,0}\Big) \otimes \overline{
    |0;0;0,0\rangle_{0,0}} 
    \\
    & \oplus |0;0;0,0\rangle_{0,0}\otimes \overline{\Big ( |2;0;0,0\rangle_{0,0} \oplus |1;1;0,0\rangle_{0,0} \oplus |1;0;\tfrac12,\tfrac12\rangle_{0,0} \Big) 
    } 
    \\
    &\oplus \Big ( |1;1;0,0\rangle_{0,0} \oplus |1;0;\tfrac12,\tfrac12\rangle_{0,0} \Big) \otimes \overline{ \Big (|1;1;0,0\rangle_{0,0} \oplus |1;0;\tfrac12,\tfrac12\rangle_{0,0} \Big )}
    \end{aligned}
\end{equation}
are the only massless states in our theory. From the first two lines we read the dilaton, the two helicities of the graviton and the spin $1$ gauge fields, while in the third line we read the massless scalars associated to the moduli of the internal space $S^3 \times T^4$.  
Contrary to the type IIB theory\footnote{In particular, we have massless fermions obtained by taking the tensor product of \eqref{eq:j1Rsector} and $|0;0;0,0\rangle_{0,0}$ providing four massless fermions and four massless gravitini. This is consistent with the realisation of the $\mathcal{N}=(4,4)$ superalgebra $\mathfrak{psu(1,1|2)} \otimes \mathfrak{psu(1,1|2)}$.}, the spectrum is purely bosonic. Moreover, the RR sector does not yield any massless bosons.
We can now describe the case the spectrally flowed representations. The mass-shell condition is then modified to be 
\begin{equation}
    \bigg ( \frac{j(1-j)}{k} + \frac{j'(1+j')}{k} + h_{T^4} +N - w \Big (m+\frac{k w}{2} \Big) -  \frac{k \, w^2}{4} -\nu \bigg ) | \psi \rangle= 0 \, ,
\end{equation}
where we recall that for $w \neq 0$ the highest weight representation of the zero mode algebra have Cartan eigenvalue $m+\frac{kw}{2}$, hence the energy and spin are given by
\begin{equation}
    E=m+\bar m+kw \, , \qquad s=|m-\bar m| \, .
\end{equation}
Therefore solving the mass-shell condition for $m=j+q$ (and for the right-moving sector $\bar m$), we obtain the dispersion relation \cite{Ferreira:2017pgt}
\begin{equation}
    \begin{aligned}
        & E=\frac12 \Big (1 - k w + \sqrt{1 - 4 k \nu  + 4 j'(1+j')  + 4 k (N+h_{T^4}) - 2 k w - 4 k q w} \Big ) + q 
        \\
        & \qquad +\frac12 \Big (1 - k w + \sqrt{1 - 4 k \nu  + 4 j'(1+j')  + 4 k (\bar N+\bar h_{T^4}) - 2 k w - 4 k q w} \Big ) + \bar q \, ,
        \\
        &s=\Big | \frac{1}{w} \big (N+h_{T^4}-\bar N-\bar h_{T^4}  \big )\Big |\, .
    \end{aligned}
\end{equation}
Although the expressions are similar for the type IIB case, the allowed values of $N$ are not the same. We can take, for instance, the NS-NS sector $\nu=\bar \nu=\frac12$, for which in the type IIB theory the only allowed states are characterised by $N \in \frac12 + \mathbb{N}$. On the other hand, for the type 0B superstring also states $N \in \mathbb{N}$ appear, as shown in fig. \ref{fig:discretespectrum}. 

For continuous representations, the situation is very different whether the spectral flow parameter vanishes or not. Indeed, for $w = 0$, these representations form unbounded representations of the zero mode algebra and give rise to tachyonic modes, whilefor $w\neq0$ the continuous representations become bounded, with highest weight provided by the Cartan eigenvalue $\alpha +\ell+\frac{kw}{2}$. In this latter situation, the highest weight is therefore real and thus cannot give rise to any tachyons. 

At $w=0$, the mass-shell condition reads
 \begin{equation}
    \frac{\frac14+p^2}{k} + N + \frac{j'(j'+1)}{k} + h_{T^4}-\nu=0 \, ,
\end{equation}
implying that 
\begin{equation}
    p=\pm \frac12 \sqrt{-1-4 j'(1 + j') -4 k h_{T^4}+ 4k\Big (\nu-N\Big )} \, .
\end{equation}
This admits a meaningful solution if $-1-4 j'(1 + j') -4 k h_{T^4}+ 4k (\frac12-N ) \geq 0$ and hence $-1+ 4k\Big (\nu-N- h_{T^4}\Big ) \geq 4 j'(1 + j') \geq 0$. However, the last inequality can only be solved for $N+h_{T^4}<\nu$. This is the case if the level-matched NS-NS vacuum survives the GSO projection.  This means that these states are present in the type 0B superstring and correspond to tachyonic instabilities but do not appear in the type IIB superstring as expected.

Finally, we can discuss what happens in the case of spectrally flowed continuous representations. The mass-shell condition becomes 
   \begin{equation}
    \frac{\frac14 + p^2}{k} -w\Big (  \alpha+\ell \Big )-\frac{k w^2}{4}+ N + \frac{j'(j'+1)}{k} + h_{T^4}-\nu=0 \, ,
\end{equation}
identifying extremal weight states for the zero mode algebra with highest weight $\alpha+\ell+\frac{kw}{2}$. As before, we can solve for such weight for $w\neq 0$ to get
\begin{equation}
    \alpha+\ell+\frac{k w}{2}= \frac{1 + 4 j' (1 + j') + 4 p^2 + 4k (N+ h_{T^4}-\nu) + k^2 w^2}{4 k w} \, .
\end{equation}
Hence the energy and spin providing the dispersion relation are given by \cite{Ferreira:2017pgt}
\begin{equation}
    \begin{aligned}
         &E=  \frac{1 + 4 j' (1 + j') + 4 p^2 - 4k\nu   + k^2 w^2}{2 k w} + \frac{1}{w} \Big ( N+h_{T^4}+\bar N +\bar h_{T^4}\Big ) \, , 
    \\
    &s= \Big | \frac{1}{w} \Big ( N+h_{T^4}-\bar N-\bar h_{T^4} \Big) \Big | \, .
    \end{aligned}
\end{equation}
As for the case of discrete representations, the dispersion relation is similar to the type IIB case but, since a different GSO projection is at play, the values of $N$ and $\bar N$ occurring are different. For the purpose of this paper, these representations do not play any role, even though they are crucial for the description of the massive spectrum and hence to identify the dual holographic CFT. Hence, we will ignore these representations in the following.

In the type IIB superstring, it was observed in \cite{Eberhardt:2017pty} that the spectrum described in \cite{Ferreira:2017pgt} and reproduced here seem to break the expected T-duality of the theory. Nonetheless, the missing states are expected to be hidden in the instanton singularity \cite{Seiberg:1999xz}. It would be interesting to see how T-duality is restored also in this non-supersymmetric set-up.

\subsection{AdS\texorpdfstring{$_3 \times S^3 \times S^3 \times S^1$}{3 x S3 x S3 x S1}}

The quantization of the type 0B superstring for the AdS$_3 \times S^3 \times S^3 \times S^1$ backgrounds follows the same philosophy. In this case however the criticality condition reads
\begin{equation}\label{eq:criticalityS3S1}
    \frac{3(k+2)}{k} + \frac{3(k_1-2)}{k_1} + \frac{3(k_2-2)}{k_2} + \frac{3}{2} + \frac{9}{2} =15 \, ,
\end{equation}
thus implying $k= k_1 k_2 /(k_1+k_2)$\footnote{Notice that the tensionless limit in this case is described by $k=1$ hence $k_1=k_2=2$ which accommodates for a RNS superstring description as well.}. The partition function for such a background then is similar to the one described in \eqref{eq:toruspartitionfunctionT^4} and reads
\begin{equation}
\begin{aligned} \label{eq:bosonsS3S1}
    \mathcal{T}_{0\text{B}}&(S^3\times S^1)
    \\
    &= \int_{\mathcal{F}} d \mu \frac{e^{- \pi k \frac{y_2^2}{\tau_2}-4 \pi k u_2}}{\sqrt{\tau_2}} \bigg \{  \frac{e^{2 \pi \frac{y_2^2}{\tau_2}} }{\big | \Jacobitheta{1/2}{1/2} \big |^2}  + \frac{\sum_{w ,\ell}\delta_{w,\ell} }{\big | \eta \big |^6} \bigg \}\sum_{\ell_1=0}^{\frac{k_1-2}{2}} |\chi_{\ell_1}|^2 \sum_{\ell_2=0}^{\frac{k_2-2}{2}} |\chi_{\ell_2}|^2 \ \Gamma_{(1,1)} \big ( \eta \bar \eta \big )^2
     \\
     & \qquad \qquad \times \, \frac{e^{-8 \pi u_2}}{2}\sum_{a,b=0,\frac12} \frac{\Jacobitheta{a}{b}^4}{ \eta^4}\frac{\Jacobibartheta{a}{b}^4}{ \bar \eta^4} 
     \\
     &=\int_{\mathcal{F}} d \mu \, B(\tau,\bar \tau \, | \, S^3\times S^1) F_{0\text{B}}(\tau,\bar \tau \,| \,S^3\times S^1) \, ,
\end{aligned}
\end{equation}
As before, we have included into $B(\tau,\bar \tau \, | \, S^3\times S^1)$ the contributions coming from the world-sheet bosons, the two $\widehat{\mathfrak{su(2)}}_k$ diagonal modular invariants, the Narain lattice depending only on the internal radius $R$ as $\Gamma_{(1,1)}(R)$ and  the $(b,c)_{(2,-1)}$ system which compensate the time-like direction in AdS$_3$ and the internal $S^1$ \cite{Evans:1998qu}. The contribution from the world-sheet fermions is conveniently expressed in terms of the following characters
\begin{equation}
\begin{aligned}
    F_{0\text{B}}(\tau, \bar \tau \, | \, S^3 \times S^1)=  \, e^{-8 \pi u_2} &\Big \{   \big | O_2  O_3  O_3 + V_2  V_3  O_3  + V_2  O_3  V_3  + O_2  V_3  V_3   \big |^2
        \\
        & \quad  + \big | O_2  O_3  V_3 + V_2  V_3  V_3  + V_2  O_3  O_3   + O_2  V_3  O_3  \big |^2
        \\
        &\quad + \big | S_2  S_3  S_3  +  C_2  S_3  S_3  \big |^2  + \big | C_2  S_3  S_3  +  S_2  S_3  S_3  \big |^2 \Big \} \, ,
    \end{aligned}
    \end{equation}
where we have omitted the dependence on the chemical potential $z=e^{2 \pi i y}$ for the $\widehat{\mathfrak{so(2)}}_1$ characters, while the two $\widehat{\mathfrak{so(3)}}_1$ characters refer to the internal $S^3$ spaces. As in the previous case it is convenient to keep track of the $\mathfrak{su(2)}$ representations by introducing the chemical potentials $z_j=e^{2 \pi i y_j}$, with $j=1,2$\footnote{As discussed in the previous Section, this also implies introducing the chemical potenitals for the central element of the affine algebra $e^{2 \pi i u_j}$ to preserve modular invariance.}. Again, we have split the expression into two factors which are independently modular invariant. The spectrum can be described in a similar way by looking at those states that satisfy the level-matching condition appearing at a given level $N$ with spins $j,j_1,j_2$. Hence once these data are known the value of $j$ characterising the representation is read in \eqref{eq:massshellT4} by shifting $\mathcal{Q_{\mathfrak{su(2)}}}/k \to \mathcal{Q}^{(1)}_{\mathfrak{su(2)}}/k_1 + \mathcal{Q}^{(2)}_{\mathfrak{su(2)}}/k_2 $ and $h_{T^4} \to h_{S^1}$. We then evaluate the spectrum by focusing on the low-energy theory which then requires $h_{S^1}=0$.

The discrete unflowed representations require both for R ground states and $N=\frac12$ states in the NS sector
\begin{equation}
    \frac{j(1-j)}{k} +  \frac{j_1(1-j_1)}{k_1} +  \frac{j_2(1-j_2)}{k_2}=0 \, ,
\end{equation}
which is solved if $j=j_1+1=j_2+1$ once the criticality condition \eqref{eq:criticalityS3S1} is imposed. 
The NS $N=\frac12$ states are obtained by looking at the expansion 
    \begin{equation}
    \begin{aligned}\label{eq:V8charactersS3S1}
        &V_2\Big ( V_3 V_3+ O_3 O_3\Big ) \sim ((z^{-1}+z) q^{\frac12}+\ldots) \Big [  1+\ldots \Big ] \, ,
        \\
        &O_2 \Big ( O_3 V_3 + V_3 O_3 \Big )
          \sim \big ( 1 +\ldots \big ) \Big [ (1+ \ldots )(z_2^{-1} +1 +z_2)q^{\frac12} +\ldots)
          \\
          & \qquad \qquad \qquad \qquad \qquad \qquad \qquad + ((z_1^{-1}+1+z_1) q^{\frac12} +\ldots )(1 +\ldots) \Big ] \, .
    \end{aligned}
    \end{equation}
from which we can read the low-energy spectrum when combined with the states $|j;j-1;j-1\rangle_{0,0}$. This gives 
\begin{equation}
\begin{aligned}
     &|j-1;j-1;j-1\rangle_{0,0} \oplus |j+1;j-1;j-1 \rangle_{0,0} \oplus
    |j;j-2;j-1\rangle_{0,0}  \oplus |j;j-1;j-1 \rangle_{0,0}
    \\
    &\quad\oplus |j;j;j-1 \rangle_{0,0} \oplus |j;j-1;j-2\rangle_{0,0} \oplus |j;j-1;j-1 \rangle_{0,0} \oplus |j;j-1;j \rangle_{0,0} \, ,
    \end{aligned}
\end{equation}
where we have denoted a generic state with $\mathfrak{sl}(\mathfrak{2},\mathbb{R})$ spin $j$, $\mathfrak{su(2)}$ spin $j_1$, $\mathfrak{su(2)}$ spin $j_2$, KK momenta and winding $m,n$ as $|j;j_1;j_2\rangle_{m,n}$. 
As in the previous Section, the $j=1$ case is non-trivial since the $D_0^+$ is reducible. In particular, the $j=1, j_1=j_2=0$ case gives
\begin{equation}
        |0;0;0\rangle_{0,0}  \oplus  |1;0;0\rangle_{0,0}  \oplus |2;0;0\rangle_{0,0} \oplus |1;0;1\rangle_{0,0} \oplus |1;1;0\rangle_{0,0} \, .
\end{equation}
From the R sector, the world-sheet oscillators provide
 \begin{equation}
    \begin{aligned}\label{eq:RsectorS3S1}
        &S_2 S_3 S_3 \sim (z^{\frac12} q^{\frac18}+\ldots)  ((z_1^{\frac12} + z_1^{-\frac12} ) q^{\frac{3}{16}}+\ldots)((z_2^{\frac12} + z_2^{-\frac12} ) q^{\frac{3}{16}}+\ldots) \, ,
        \\
        &C_2 S_3 S_3\sim (z^{-\frac12} q^{\frac18}+\ldots)  ((z_1^{\frac12} + z_1^{-\frac12} ) q^{\frac{3}{16}}+\ldots)((z_2^{\frac12} + z_2^{-\frac12} ) q^{\frac{3}{16}}+\ldots) \, .
    \end{aligned}
    \end{equation}
Hence acting on the ground states $|j;j-1;j-1\rangle_{0,0}$ they provide the following contribution
\begin{equation}
    \begin{aligned}
        &|j+\tfrac12;j-\tfrac32;j-1\rangle_{0,0} \oplus |j+\tfrac12;j-\tfrac12;j-1 \rangle_{0,0} \oplus |j+\tfrac12;j-1;j-\tfrac32 \rangle_{0,0}         
        \\
        &\oplus |j +\tfrac12;j-1;j-\tfrac12\rangle_{0,0} \oplus |j-\tfrac12;j-\tfrac32;j-1\rangle_{0,0} \oplus |j-\tfrac12;j-\tfrac12;j-1 \rangle_{0,0} 
        \\
        &\oplus |j-\tfrac12;j-1;j-\tfrac32 \rangle_{0,0} \oplus |j -\tfrac12;j-1;j-\tfrac12\rangle_{0,0} \, .
    \end{aligned}
\end{equation}
In the $j=1, j_1=0,j_2=0$ case, we obtain
\begin{equation}
        |\tfrac32;\tfrac12;\tfrac12\rangle_{0,0}  \oplus    |\tfrac12;\tfrac12;\tfrac12\rangle_{0,0} \, .
\end{equation}
Summing these two sectors according to the GSO projection of the type IIB superstring \eqref{eq:GSOIIB} we obtain 
\begin{equation}
\begin{aligned}
    F_{II\text{B}}(\tau, \bar \tau | S^3 \times S^1)=   \, e^{-8 \pi u_2}   & \big | O_2  O_3  V_3 + V_2  V_3  V_3  + V_2  O_3  O_3  
        \\
        &  + O_2  V_3  O_3  - S_2  S_3  S_3  -  C_2  S_3  S_3  \big |^2 \, ,
    \end{aligned}
\end{equation}
whose light spectrum is organised as
\begin{equation}
\begin{aligned}
    \bigoplus_{j \in I'} \Big ( &(j-1;j-1;j-1)_{s} \oplus (j-\tfrac12;j-\tfrac12;j-\tfrac12)_{s} \Big ) 
    \\
    &\otimes \overline{\Big ( (j-1;j-1;j-1)_{s} \oplus (j-\tfrac12;j-\tfrac12;j-\tfrac12)_{s} \Big )} \, ,
    \end{aligned}
\end{equation}
where $I'=\{ j  \in \mathbb{Z}/2 \, , \, j\geq 1, \, j \leq \text{min}(\frac{k+1}{2},\frac{k_1}{2},\frac{k_2}{2})]\}$.
Here $(j)_s$ denotes a short representation for the $D(2,1|\alpha)$ superalgebra possessing a {\em large} $\mathcal{N}=4$ supersymmetry. It is decomposed as  \begin{equation}
\begin{aligned}
 (j;j_1;j_2)_s=&|j;j_1;j_2\rangle \oplus | j+\tfrac12;j_1+\tfrac12;j_2-\tfrac12\rangle \oplus |j+\tfrac12;j_1-\tfrac12;j_2+\tfrac12\rangle 
 \\
 &\oplus |j+\tfrac12;j_1-\tfrac12;j_2-\tfrac12\rangle \oplus |j+1;j_1-1;j_2\rangle \oplus |j+1;j_1;j_2\rangle 
 \\
 &\oplus |j +1;j_1;j_2-1\rangle\oplus |j+\tfrac32;j_1-\tfrac12;j_2-\tfrac12\rangle \, ,
 \end{aligned}
\end{equation}
where have expressed the representations of the $\mathfrak{sl}(\mathfrak{2},\mathbb{R}) \oplus \mathfrak{su(2)} \oplus \mathfrak{su(2)}$ algebra as $|j;j_1,j_2\rangle$. For the $j=1$ case the spectrum reads
\begin{equation}
    \Big ( (0,0,0)_s \oplus (\tfrac12,\tfrac12,\tfrac12)_s \Big ) \otimes \overline{ \Big ( (0,0,0)_s \oplus (\tfrac12,\tfrac12,\tfrac12)_s \Big )} \, ,
\end{equation}
where $(0,0,0)_s=|0;0;0\rangle$, while $(\frac12;\frac12;\frac12)_s=|\frac12;\frac12;\frac12\rangle \oplus |1;1,0\rangle \oplus |1;0;1\rangle \oplus |1;0;0\rangle \oplus  |\frac32; \frac12;\frac12 \rangle \oplus |2;0;0\rangle$. The spectrum thus reproduces the result of \cite{Eberhardt:2017fsi,Eberhardt:2017pty}, matching the correct supergravity description of \cite{Eberhardt:2017pty}.

However, for the type 0B superstring we have additional sectors to take into account and a different way to combine them. In particular from the NS sector we also have 
\begin{equation}
    \begin{aligned} \label{eq:O8charactersS3S1}
        &V_2\Big ( V_3 O_3+ O_3 V_3\Big ) \sim ((z^{-1}+z) q^{\frac12}+\ldots) \Big [ ((z_1^{-1}+1+z_1) q^{\frac12} +\ldots )(1 +\ldots)
        \\
          & \qquad \qquad \qquad \qquad \qquad \qquad \qquad (1+ \ldots )(z_2^{-1} +1 +z_2)q^{\frac12}  \Big ] \, ,
        \\
        &O_2 \Big ( O_3 O_3 + V_3 V_3 \Big )
          \sim \big ( 1 +\ldots \big ) \Big [ (1+ \ldots )(1 +\ldots)
          \\
          & \qquad \qquad \qquad \qquad \qquad \qquad \qquad + ((z_1^{-1}+1+z_1) q^{\frac12} +\ldots )(z_2^{-1} +1 +z_2)q^{\frac12} +\ldots)\Big ] \, .
    \end{aligned}
\end{equation}
One can notice that we have no states at $N=\frac12$ but only at $N=0$. The level matching condition then implies that 
\begin{equation}
    j=\frac12+\sqrt{\frac{\Delta(j_1,j_2)}{4(k_1+k_2)}}   \, ,
\end{equation}
where $\Delta(j_1,j_2)= (1+2j_1)k_1+(1+2j_2)k_2-2k_1 k_2$. 
In the type 0B superstring these states are level-matched thus contributing to the spectrum as
\begin{equation}
    \bigoplus_{\substack{\frac12 <j<\frac{k+1}{2} \\ \Delta(j_1,j_2) \geq 0}} |j;j_1;j_2\rangle \otimes  \overline{{|j;j_1;j_2\rangle}} \, .
\end{equation}  
As before these states depend on the AdS curvature and thus contribute to the massive spectrum. The other contribution from the R sector is identical to the one computed in \eqref{eq:RsectorS3S1}. Therefore, once the GSO projection in \eqref{eq:0BGSOprojection} is implemented we obtain the low-energy spectrum 
\begin{equation}
   \begin{aligned}
     &\bigoplus_{j \in I'} \Big (   |j-1;j-1;j-1\rangle_{0,0} \oplus |j+1;j-1;j-1 \rangle_{0,0} \oplus
    |j;j-2;j-1\rangle_{0,0} 
    \\
    &\qquad \oplus |j;j-1;j-1 \rangle_{0,0} \oplus |j;j;j-1 \rangle_{0,0} \oplus |j;j-1;j-2\rangle_{0,0} \oplus |j;j-1;j-1 \rangle_{0,0} \oplus |j;j-1;j \rangle_{0,0} \Big )
    \\
    & \quad \otimes \overline{\Big (   |j-1;j-1;j-1\rangle_{0,0} \oplus |j+1;j-1;j-1 \rangle_{0,0} \oplus
    |j;j-2;j-1\rangle_{0,0}} 
    \\
    &\qquad  \overline{\oplus |j;j-1;j-1 \rangle_{0,0} \oplus |j;j;j-1 \rangle_{0,0} \oplus |j;j-1;j-2\rangle_{0,0} \oplus |j;j-1;j-1 \rangle_{0,0} \oplus |j;j-1;j \rangle_{0,0} \Big )} 
    \\
      &  \bigoplus_{j \in I'} 2 \Big (|j+\tfrac12;j-\tfrac32;j-1\rangle_{0,0} \oplus |j+\tfrac12;j-\tfrac12;j-1 \rangle_{0,0} \oplus |j+\tfrac12;j-1;j-\tfrac32 \rangle_{0,0}  
      \\
      & \qquad \oplus |j +\tfrac12;j-1;j-\tfrac12\rangle_{0,0} \oplus |j-\tfrac12;j-\tfrac32;j-1\rangle_{0,0} \oplus |j-\tfrac12;j-\tfrac12;j-1 \rangle_{0,0} 
      \\
      &\qquad \oplus |j-\tfrac12;j-1;j-\tfrac32 \rangle_{0,0} \oplus |j -\tfrac12;j-1;j-\tfrac12\rangle_{0,0} \Big ) \otimes \overline{\Big (|j+\tfrac12;j-\tfrac32;j-1\rangle_{0,0}}
      \\
      &\qquad \overline{\oplus |j+\tfrac12;j-\tfrac12;j-1 \rangle_{0,0} \oplus |j+\tfrac12;j-1;j-\tfrac32 \rangle_{0,0} \oplus |j +\tfrac12;j-1;j-\tfrac12\rangle_{0,0}}
        \\
        &\qquad  \overline{\oplus |j-\tfrac12;j-\tfrac32;j-1\rangle_{0,0} \oplus |j-\tfrac12;j-\tfrac12;j-1 \rangle_{0,0} \oplus |j-\tfrac12;j-1;j-\tfrac32 \rangle_{0,0} }
        \\
        &\qquad  \overline{\oplus |j -\tfrac12;j-1;j-\tfrac12\rangle_{0,0} \Big )} \, .
    \end{aligned}
\end{equation}
As in the previous case we can read the massless spectrum by looking at $j=1$.
We obtain 
\begin{equation}
\begin{aligned}
    &\Big (|0;0;0\rangle_{0,0}  \oplus  |1;0;0\rangle_{0,0}  \oplus |2;0;0\rangle_{0,0} \oplus |1;0;1\rangle_{0,0} \oplus |1;1;0\rangle_{0,0} \Big) \otimes \overline{|0;0;0\rangle_{0,0} }
    \\
    & \quad  \oplus |0;0;0\rangle_{0,0} \otimes \overline{\Big ( |1;0;0\rangle_{0,0}  \oplus |2;0;0\rangle_{0,0} \oplus |1;0;1\rangle_{0,0} \oplus |1;1;0\rangle_{0,0} \Big)} 
    \\
    &\quad  \oplus \Big (|1;0;0\rangle_{0,0}  \oplus |1;0;1\rangle_{0,0} \oplus |1;1;0\rangle_{0,0} \Big ) \otimes \overline{\Big ( |1;0;0\rangle_{0,0}  \oplus |1;0;1\rangle_{0,0} \oplus |1;1;0\rangle_{0,0} \Big)}  \, .
\end{aligned}
\end{equation}
Hence, from the expression above we read again the dilaton, the two helicities of the graviton, the gauge fields transforming in the adjoint of $\mathfrak{su(2)}\oplus \mathfrak{su(2)}$, while in the last line we read the massless scalars associated with the moduli of the internal space. The RR sector instead only contributes with massive states.

As before, the $N=0$ states are level-matched implying tachyons to be present from continuous representations. Indeed, in the unflowed case the mass shell condition reads
  \begin{equation}
    \frac{\frac14+p^2}{k} + N + \frac{j_1(j_1+1)}{k_1} + \frac{j_2(j_2+1)}{k_1} + h_{S^1}-\nu=0 \, ,
\end{equation}
which admits as a solution 
\begin{equation}\label{eq:tachyonsS3S1}
    p=\pm \frac12 \sqrt{-1-4k j_1(1 + j_1)/k_1 -4k j_2(1 + j_2)/k_2 + 4k\Big (\nu-N-h_{S^1}\Big )} \, .
\end{equation}
One can show that this is a real expression only if $-1+4k(\nu-N-h_{S^1})\geq 0$ for which only $N=0$ is allowed. Therefore, as for the $T^4$ case, these states are absent in the type IIB superstring but are allowed for the type 0B, reflecting the instability of such model. 

\section{The heterotic string} \label{sec:16x16}

The discussion for the heterotic string follows a similar pattern \cite{Kutasov:1998zh, Hohenegger:2008du}. In this case however, the world-sheet fermions are present only in the holomorphic sector, while the anti-holomorphic sector is composed by compact bosons decoupled from the current of the WZW model. 
The world-sheet theory is described by a $\widehat{\mathfrak{sl}(\mathfrak{2},\mathbb{R})}_{k_s+2}$ Ka$\check{\text{c}}$-Moody algebra in the holomorphic sector, with $k_s=k-2$ the level of the algebra with world-sheet fermions decoupled, while in the anti-holomorphic sector we still have the affine algebra realised at level $k$\footnote{Notice that the level of the purely bosonic $\mathfrak{sl(2,} \mathbb{R}\mathfrak{)}$ algebras in the holomorphic and anti-holomorphic sectors are the same.}.
In the following, we are going to adapt the discussion presented for the superstring theory to the heterotic case by describing AdS$_3$ backgrounds. We will focus on the non-supersymmetric $\text{Spin}(16) \times \text{Spin}(16) \rtimes \mathbb{Z}_2$ theory although for the sake of completeness we report the partition function and low-energy spectrum of the supersymmetric $ \text{E}_8 \times \text{E}_8 \rtimes \mathbb{Z}_2$ theory in the Appendix \ref{app:SUSYE8}.

\subsection{The tachyon-free \texorpdfstring{$\text{Spin}(16) \times\text{Spin}(16) \rtimes \mathbb{Z}_2$}{ Spin(16)x Spin(16)x Z2} heterotic string}

There are many ways through which the $\text{Spin}(16) \times \text{Spin}(16) \rtimes \mathbb{Z}_2$ heterotic theory can be described. For instance, one can start from a supersymmetric theory, either the $\text{Spin}(32)/\mathbb{Z}_2$ or the $ \text{E}_8 \times \text{E}_8 \rtimes \mathbb{Z}_2$ theory, and perform a freely-acting orbifold by flipping the sign of space-time fermions and the spinors of $\text{Spin}(16) \subset E_8$ or $\text{Spin}(32)$. The resulting GSO projection gives rise to a non-supersymmetric theory without level-matched $N=0$ states. This implies in flat space, and, as we  will see, in AdS$_3$ spaces as well, that tachyons are absent from the spectrum. However, in flat space geometric compactifications, it is known that marginal deformations can be turned on to destabilise the theory \cite{Ginsparg:1986wr}. In this Section, we show that a similar situation holds for AdS$_3$ spaces with a concrete example. 

\subsubsection{AdS\texorpdfstring{$_3 \times S^3 \times T^4$}{3 x S3 x T4}}

The criticality condition for the holomorphic sector is the same as the one described for the superstring in \eqref{eq:criticalitytype0T4}, thus implying $k_s=k_s'$ (or equivalently $k=k'+4$), with $k_s=k-2$ and $k_s'=k'+2$. This allows to solve the condition on the anti-holomorphic sector 
\begin{equation}
    \frac{3 (k_s+2)}{k_s}+ \frac{3 (k_s-2)}{k_s} + 4 + 16=26 \, .
\end{equation}
The partition function describing this theory is then
\begin{equation}\label{eq:toruspartitionfunctionT^416x16}
\begin{aligned}
    \mathcal{T}_{16 \times 16}(T^4)&= \int_{\mathcal{F}} d \mu \frac{e^{- \pi k \frac{y_2^2}{\tau_2}-4 \pi k u_2}}{\sqrt{\tau_2}} \bigg \{  \frac{e^{2 \pi \frac{y_2^2}{\tau_2}} }{\big | \Jacobitheta{1/2}{1/2} \big |^2}  + \frac{\sum_{w ,\ell}\delta_{w,\ell} }{\big | \eta \big |^6} \bigg \}\sum_{\ell=0}^{\frac{k-2}{2}} |\chi_{\ell}|^2 \ \Gamma_{(4,4)} \big ( \eta \bar \eta \big )^2
     \\
     & \   \times \, \frac{e^{-4 \pi i u}}{4 \eta^4 \bar \eta^{16}} \Big [ \Jacobitheta{0}{0}^4 \Jacobibartheta{0}{1/2}^8\big ( \Jacobibartheta{0}{0}^8-\Jacobibartheta{1/2}{0}^8 \big ) - \Jacobitheta{0}{1/2}^4 \Jacobibartheta{0}{0}^8\big ( \Jacobibartheta{0}{1/2}^8-\Jacobibartheta{1/2}{0}^8 \big ) \Big ]
     \\
     &= \int_{\mathcal{F}} d \mu \, B_H(\tau,\bar \tau \, | \, T^4) F_{16 \times 16}(\tau,\bar \tau \,| \,T^4) \, ,
    \end{aligned}
\end{equation}
where as before we have organised the expression into two independent modular invariant pieces. Notice that the bosonic contribution $B_H(\tau,\bar \tau|T^4)$ coincides with \eqref{eq:toruspartitionfunctionT^4} if we replace $k_s$ with $k$. This difference originates from different levels of the affine algebra occurring in left and right moving sectors $(k_s,k)$. Indeed, the fermionic contribution which enters only the left moving sector is responsible for the appearance of $k_s$, while in the right moving sector is purely bosonic so that the level entering in the partition function is the same as the bosonic theory \eqref{eq:bosAdS3}. Since $k_s=k-2$, then the result in \eqref{eq:toruspartitionfunctionT^416x16} is obtained. 
The contribution coming from the world-sheet left-moving fermions and the compact bosons in the anti-holomorphic sector can be conveniently reorganised as  \cite{Alvarez-Gaume:1986ghj, Dixon:1986iz}
\begin{equation}
    \begin{aligned}
         F_{16 \times 16}(\tau,\bar \tau \,| \,T^4)= e^{-4 \pi i u} &\bigg \{\Big ( V_2 O_2 O_4 + V_2 V_2 V_4 + O_2 V_2 O_4+ O_2 O_2 V_4 \Big) \Big ( \bar O_{16} \bar O_{16} + \bar S_{16} \bar S_{16} \Big )
  \\
  & + \Big ( O_2  O_2 O_4 + O_2 V_2 V_4 + V_2 V_2 O_4+ V_2 O_2 V_4 \Big )  \Big ( \bar V_{16} \bar C_{16} + \bar C_{16} \bar V_{16} \Big )
        \\
        &  - \Big (S_2 S_2 S_4 + S_2 C_2 C_4 + C_2 S_2 C_4 + C_2 C_2 S_4 \Big ) \Big ( \bar O_{16} \bar S_{16} + \bar S_{16} \bar O_{16} \Big )
        \\
        & - \Big (C_2 S_2 S_4 + C_2 C_2 C_4 + S_2 S_2 C_4 + S_2 C_2 S_4 \Big ) \Big ( \bar V_{16} \bar V_{16} + \bar C_{16} \bar C_{16} \Big ) \bigg \} \, .
    \end{aligned}
\end{equation}
As in the previous case, the first set of $\widehat{\mathfrak{so(2)}}_1$ characters are associated with the $\mathfrak{sl}(\mathfrak{2},\mathbb{R})$ world-sheet fermions and carry an implicit dependence on the chemical potential $z=e^{2\pi i y}$ and the second set refers to the $\mathfrak{su(2)}$ ones. 
Notice that now the factor depending on $u$ is different with respect to the type 0B superstring and it is crucial to guarantee modular invariance: this arises naturally from the difference $2 \pi i u k_s - 2 \pi i \bar u k$ and compensates the modular transformation of the holomorphic world-sheet fermions.

Following the same steps as in Section \ref{sec:type0}, we can turn on a chemical potential for the $\mathfrak{su(2)}$ algebra as well $z'=e^{2\pi i y'}$ and read the low-energy spectrum by looking at the unflowed sectors. For the discrete case the lowest levels give again rise to the condition \eqref{eq:conditionunflowedT4}. 
Hence for these representations $j=j'+1$ as before. The holomorphic sector can be then directly read from the previous expansions in \eqref{eq:V8expansionT4}, \eqref{eq:S8expansionT4}, \eqref{eq:O8expansionT4} and \eqref{eq:C8expansionT4}, while the anti-holomorphic sector now encodes the gauge part. In particular, a straightforward computation gives 
\begin{equation}
    \begin{aligned}\label{eq:so16characters}
        & \Big ( \bar O_{16} \bar O_{16} + \bar S_{16} \bar S_{16} \Big ) \sim 1 + (120+120) \bar q + \ldots  \, ,
  \\
  &   \Big ( \bar V_{16} \bar C_{16} + \bar C_{16} \bar V_{16} \Big ) \sim (16 \cdot 128+ 16 \cdot 128) \bar q^{\frac32} + \ldots \, ,
        \\
        &  \Big ( \bar O_{16} \bar S_{16} + \bar S_{16} \bar O_{16} \Big ) \sim (128+128) \bar q + \ldots \, ,
        \\
        &  \Big ( \bar V_{16} \bar V_{16} + \bar C_{16} \bar C_{16} \Big ) \sim 16 \cdot 16 \, \bar q + \ldots \, ,
    \end{aligned}
\end{equation}
where we have omitted $\bar q^{-\frac{16}{24}}$ in front of each expansion. Hence we can now read the spectrum, once we impose the level-matching condition. This immediately implies that there are no $N=0$ level matched states. Hence the low-energy spectrum reads 
\begin{equation}
\begin{aligned}
    &\bigoplus_{j \in I}\Big ( |j-1;j-1;0,0\rangle_{0,0} \oplus |j+1;j-1;0,0\rangle_{0,0} \oplus |j;j;0,0\rangle_{0,0} \oplus |j;j-2;0,0\rangle_{0,0} 
    \\
    &\qquad \qquad \oplus |j;j-1;\tfrac12,\tfrac12\rangle_{0,0} \Big ) \otimes \overline{\Big ( |j-1;j-1;0,0; \boldsymbol{1};\boldsymbol{1}\rangle_{0,0} \oplus |j+1;j-1;0,0; \boldsymbol{1};\boldsymbol{1}\rangle_{0,0} }
    \\
    &\qquad \qquad \qquad \overline{\oplus |j;j;0,0; \boldsymbol{1};\boldsymbol{1}\rangle_{0,0} \oplus |j;j-2;0,0; \boldsymbol{1};\boldsymbol{1}\rangle_{0,0}\oplus |j;j-1;\tfrac12,\tfrac12; \boldsymbol{1};\boldsymbol{1}\rangle_{0,0}}
    \\
    &\qquad \qquad \qquad\overline{\oplus|j;j-1;0,0; \boldsymbol{120};\boldsymbol{1}\rangle_{0,0} \oplus |j;j-1;0,0; \boldsymbol{1};\boldsymbol{120}\rangle_{0,0} \Big )  }
    \\
    &\qquad \qquad \oplus \Big (|j+\tfrac12;j-\tfrac12;\tfrac12,0\rangle_{0,0} \oplus  |j+\tfrac12;j-\tfrac32;0,\tfrac12\rangle_{0,0} \oplus  |j-\tfrac12;j-\tfrac12;0,\tfrac12\rangle_{0,0}
    \\
    & \qquad \qquad \oplus |j-\tfrac12;j-\tfrac32;\tfrac12,0\rangle_{0,0} \Big ) \otimes \overline{\Big ( |j;j-1;0,0; \boldsymbol{128};\boldsymbol{1}\rangle_{0,0} \oplus |j;j-1;0,0; \boldsymbol{1};\boldsymbol{128}\rangle_{0,0} \Big )  }
    \\
    &  \qquad \qquad \oplus \Big ( |j+\tfrac12;j-\tfrac12;0,\tfrac12\rangle_{0,0} \oplus  |j+\tfrac12;j-\tfrac32;\tfrac12,0\rangle_{0,0} \oplus  |j-\tfrac12;j-\tfrac12;\tfrac12,0\rangle_{0,0}
    \\
    & \qquad \qquad  \oplus |j-\tfrac12;j-\tfrac32;0,\tfrac12\rangle_{0,0} \Big ) \otimes \overline{\Big ( |j;j-1;0,0; \boldsymbol{16};\boldsymbol{16}\rangle_{0,0}  \Big )  } \, ,
    \end{aligned}
\end{equation}
where here $I=\{ j  \in \mathbb{Z}/2 \, , \, j\geq 1, \, j \leq \frac{k_s}{2} \}$.
In the expression above we have considered the additional contribution coming from the $N=1$ bosonic oscillator, which are present since the zero point energy of the anti-holomorphic sector is $\nu=1$. For instance for the level-matched representations 
\begin{equation}
\begin{aligned} \label{eq:antiholobosT4}
     \bar \chi^{+}_{j,0} \  \bar \chi_{j-1} \ \bar \eta^{-4} \ \bar \eta^2 \sim &  \Big ( \bar z+ \bar z^2 + \bar z^3 +\ldots + ( 1+   \bar z + 2 \bar z^2+\ldots ) \bar q+ O(\bar q^2) \Big ) 
     \\
     &\Big ( 1+ (\bar z'^{-1} + \bar z') \bar q + O(\bar q^2)\Big ) \Big ( 1+4 \bar q+ O(\bar q^2) \Big ) \, ,
\end{aligned}
\end{equation}
where we have omitted the overall contribution to the zero point energy. As expected, from the $\bar q^{0}$ row of the $\bar \chi^+_{j,0}$ character we can read the $D_j^+$ representation while from the $\bar q$ piece we can read $D_{j-1}^+ \oplus D_{j+1}^+ $ representation. 
When dealing with the $j=1$, we have to take into account that $D_0^+$ is not irreducible and decomposes as \eqref{eq:j0representationsdecomposition}. Hence, the $j=1$ the contribution in \eqref{eq:antiholobosT4} gives
\begin{equation}
   \big ( \mathbf{1} \otimes  \boldsymbol{1}  \otimes  \boldsymbol{1} \big ) \oplus \big ( D_1^+ \otimes  \boldsymbol{3} \otimes  \boldsymbol{1} \big ) \oplus \big ( D_1^+ \otimes  \boldsymbol{1} \otimes  \boldsymbol{4} \big ) \oplus \big ( D_2^+ \otimes  \boldsymbol{1} \otimes  \boldsymbol{1} \big ) \, .
\end{equation}
Adding the contribution from the compact bosons encoding the $\text{Spin}(16) \times \text{Spin}(16) \rtimes \mathbb{Z}_2$ gauge group, the $j=1$ NS spectrum is given by 
\begin{equation}
\begin{aligned}
        &\Big (|0;0;0,0\rangle_{0,0} \oplus |2;0;0,0\rangle_{0,0} \oplus |1;1;0,0\rangle_{0,0} \oplus |1;0;\tfrac12,\tfrac12\rangle_{0,0}\Big)
        \\
        &\otimes \overline{\Big (|0;0;0,0;\boldsymbol{1};\boldsymbol{1}\rangle_{0,0} \oplus |2;0;0,0;\boldsymbol{1};\boldsymbol{1}\rangle_{0,0} \oplus |1;1;0,0;\boldsymbol{1};\boldsymbol{1}\rangle_{0,0} \oplus |1;0;\tfrac12,\tfrac12;\boldsymbol{1};\boldsymbol{1}\rangle_{0,0}}
        \\
        & \qquad \overline{\oplus |1;0;0,0;\boldsymbol{120};\boldsymbol{1}\rangle_{0,0} \oplus |1;0;0,0;\boldsymbol{1};\boldsymbol{120}\rangle_{0,0} \Big)}\, .
        \end{aligned}
\end{equation}
The contribution coming from the R ground states is similar to the $j\neq 1$ case and reads
\begin{equation}
\begin{aligned}
    &\Big (|\tfrac12;\tfrac12,\tfrac12;0,\tfrac12\rangle_{0,0} \oplus  |\tfrac12;\tfrac12,-\tfrac12;\tfrac12,0\rangle_{0,0} \oplus  |\tfrac32;\tfrac12,\tfrac12;0,\tfrac12\rangle_{0,0} \oplus  |\tfrac32;\tfrac12,-\tfrac12;\tfrac12,0\rangle_{0,0} \Big)
    \\
    &\qquad \otimes \overline{\Big ( |1;0;0,0; \boldsymbol{128};\boldsymbol{1}\rangle_{0,0} \oplus |1;0;0,0; \boldsymbol{1};\boldsymbol{128}\rangle_{0,0} \Big )  }
    \\
    &\oplus \Big (|\tfrac12;\tfrac12,\tfrac12;\tfrac12,0\rangle_{0,0} \oplus  |\tfrac12;\tfrac12,-\tfrac12;0,\tfrac12\rangle_{0,0} \oplus  |\tfrac32;\tfrac12,\tfrac12;\tfrac12,0\rangle_{0,0} \oplus  |\tfrac32;\tfrac12,-\tfrac12;0,\tfrac12\rangle_{0,0} \Big)
    \\
    &\qquad \otimes \overline{\Big ( |1;0;0,0; \boldsymbol{16};\boldsymbol{16}\rangle_{0,0}  \Big )  }\, .
    \end{aligned}
\end{equation}
The only contribution to the massless spectrum then arises from the NS sector and comprises the graviton, the dilaton, the gauge field transforming in the adjoint of $\text{SU}(2) \times \text{U}(1)^4 \times \text{Spin}(16) \times \text{Spin}(16)$. The massless scalars associated with deformation moduli of $S^3 \times T^4$ and in the adjoint of the $\text{Spin}(16) \times \text{Spin}(16)$ group as can be read from  
\begin{equation}
\begin{aligned}
 &\Big ( |1;1;0,0\rangle_{0,0} \oplus |1;0;\tfrac12,\tfrac12\rangle_{0,0}\Big)
        \\
        &\otimes \overline{\Big ( |1;1;0,0;\boldsymbol{1};\boldsymbol{1}\rangle_{0,0} \oplus |1;0;\tfrac12,\tfrac12;\boldsymbol{1};\boldsymbol{1}\rangle_{0,0}\oplus |1;0;0,0;\boldsymbol{120};\boldsymbol{1}\rangle_{0,0} \oplus |1;0;0,0;\boldsymbol{1};\boldsymbol{120}\rangle_{0,0} \Big)}\, .
        \end{aligned}
\end{equation}
These scalars are crucial because the directions associated wit the Cartan generators of $\text{Spin}(16) \times \text{Spin}(16)$ correspond to {\em bona fide} Wilson lines that can be used to deform the theory. This possibility will be explicitly explored in the next Section for AdS$_3 \times S^3 \times S^3 \times S^1$ backgrounds, although the same considerations apply in this setting with little modification.

In the considerations above, we have seen that $N=0$ states in the NS sector are absent because of the level-matching condition. This means that no unflowed continuous representations are present and hence the theory has no tachyons. Continuous representations appear only in the specrally flowed sector and hence contribute to the massive spectrum of the theory.  

\subsubsection{AdS\texorpdfstring{$_3 \times S^3 \times S^3 \times S^1$}{3 x S3 x S3 x S1}}

We can now turn to the discussion of the AdS$_3 \times S^3 \times S^3 \times S^1$ background, which is the main interest of this paper. As before, imposing the criticality condition on the shifted level 
\begin{equation}
    \frac{1}{k_s}=\frac{1}{k_{s}^1}+\frac{1}{k_s^2} \, ,
\end{equation}
allows to saturate the central charge also in the anti-holomorphic sector
\begin{equation}
    \frac{3 (k_s+2)}{k_s} +  \frac{3 (k_s^1-2)}{k_s^1} + \frac{3 (k_s^2-2)}{k_s^2} +1+16=26 \, .
\end{equation}
The partition function therefore becomes 
\begin{equation}\label{eq:toruspartitionfunctionS3S116x16}
\begin{aligned}
    \mathcal{T}_{16 \times 16}&(S^3\times S^1)
    \\
    &= \int_{\mathcal{F}} d \mu \frac{e^{- \pi k \frac{y_2^2}{\tau_2}-4 \pi k u_2}}{\sqrt{\tau_2}} \bigg \{  \frac{e^{2 \pi \frac{y_2^2}{\tau_2}} }{\Big | \Jacobitheta{1/2}{1/2} \Big |^2}  + \frac{\sum_{w ,\ell}\delta_{w,\ell} }{\big | \eta \big |^6} \bigg \}\sum_{\ell_1=0}^{\frac{k_1-2}{2}} |\chi_{\ell_1}|^2 \sum_{\ell_2=0}^{\frac{k_2-2}{2}} |\chi_{\ell_2}|^2 \ \Gamma_{(1,1)} \big ( \eta \bar \eta \big )^2
     \\
     & \quad   \times \, \frac{e^{-4 \pi i u}}{4 \eta^4 \bar \eta^{16}} \Big [ \Jacobitheta{0}{0}^4 \Jacobibartheta{0}{1/2}^8\big ( \Jacobibartheta{0}{0}^8-\Jacobibartheta{1/2}{0}^8 \big ) - \Jacobitheta{0}{1/2}^4 \Jacobibartheta{0}{0}^8\big ( \Jacobibartheta{0}{1/2}^8-\Jacobibartheta{1/2}{0}^8 \big ) \Big ]
     \\ 
     \\
     &= \int_{\mathcal{F}} d \mu \, B_H(\tau,\bar \tau \, | \, S^3\times S^1) F_{16 \times 16}(\tau,\bar \tau \,| \, S^3\times S^1) \, ,
    \end{aligned}
\end{equation}
where the contribution from the world-sheet bosons corresponds to \eqref{eq:bosonsS3S1} with the replacement $k \to k_s$ and $k_{1,2} \to k_s^{1,2}$. As explained in the previous Section, this is due to the different level of the affine algebra occurring on the left and right-moving sector. The contribution coming from the world-sheet fermions and the compact bosons can be conveniently arranged in 
\begin{equation}
    \begin{aligned}
         F_{16 \times 16}(\tau,\bar \tau \,| \,S^3 \times S^1)= e^{-4 \pi i u} &\bigg \{\Big ( O_2  O_3  V_3 + V_2  V_3  V_3  + V_2  O_3  O_3   + O_2  V_3  O_3  \Big) \Big ( \bar O_{16} \bar O_{16} + \bar S_{16} \bar S_{16} \Big )
  \\
  & + \Big ( O_2  O_3  O_3 + V_2  V_3  O_3  + V_2  O_3  V_3  + O_2  V_3  V_3 \Big )  \Big ( \bar V_{16} \bar C_{16} + \bar C_{16} \bar V_{16} \Big )
        \\
        &  - \Big ( S_2  S_3  S_3  +  C_2  S_3  S_3 \Big ) \Big ( \bar O_{16} \bar S_{16} + \bar S_{16} \bar O_{16} \Big )
        \\
        & - \Big (C_2  S_3  S_3  +  S_2  S_3  S_3  \Big ) \Big ( \bar V_{16} \bar V_{16} + \bar C_{16} \bar C_{16} \Big ) \bigg \}  \, ,
    \end{aligned}
\end{equation}
where we have omitted the dependence on the chemical potential $z=e^{2\pi i y}$ on the first set of $\widehat{\mathfrak{so(2)}}_1$ characters and the factor $e^{-4\pi i u}$ is crucial to guarantee modular invariance. 
In the following, we will turn chemical potentials $z_j=e^{2\pi iy_j}$, with $j=1,2$ to read the two $\mathfrak{su(2)}$ representations and, since we are interested in the stability properties of the low-energy theory, we will focus on unflowed sector of both discrete and continuous representations. 
As in the case of the $T^4$ internal manifold, there are no $N=0$ states level-matched from the NS sector, which implies that no continuous representations are present. Hence, the only contribution to the low-energy theory comes from the discrete sector of the Hilbert space to which we now turn. 
As before, the mass-shell condition implies that $j=j_1+1=j_2+1$, and the spectrum can be directly read from the $q-$expansion of the characters involved.
This has been already discussed in the previous Sections and their contribution can be directly read from eqs. \eqref{eq:V8charactersS3S1}, \eqref{eq:RsectorS3S1}, \eqref{eq:O8charactersS3S1} and \eqref{eq:so16characters}. Therefore at low-energy
\begin{equation}
\begin{aligned}
    &\bigoplus_{j \in I'}  \Big (   |j-1;j-1;j-1\rangle_{0,0} \oplus |j+1;j-1;j-1 \rangle_{0,0} \oplus |j;j-2;j-1\rangle_{0,0} \oplus |j;j-1;j-1 \rangle_{0,0} 
    \\
    &\qquad \oplus |j;j;j-1 \rangle_{0,0} \oplus |j;j-1;j-2\rangle_{0,0} \oplus |j;j-1;j-1 \rangle_{0,0} \oplus |j;j-1;j \rangle_{0,0} \Big )
    \\
    & \quad \otimes \overline{ \Big ( |j-1;j-1;j-1; \boldsymbol{1};\boldsymbol{1}\rangle_{0,0} \oplus |j+1;j-1;j-1; \boldsymbol{1};\boldsymbol{1} \rangle_{0,0} \oplus |j;j-2;j-1; \boldsymbol{1};\boldsymbol{1}\rangle_{0,0}}
    \\
    &\qquad \overline{ \oplus |j;j-1;j-1 ; \boldsymbol{1};\boldsymbol{1} \rangle_{0,0}  |j;j;j-1 ; \boldsymbol{1};\boldsymbol{1}\rangle_{0,0} \oplus |j;j-1;j-2; \boldsymbol{1};\boldsymbol{1}\rangle_{0,0} \oplus |j;j-1;j-1; \boldsymbol{1};\boldsymbol{1} \rangle_{0,0}}
    \\
    & \qquad  \overline{\oplus |j;j-1;j; \boldsymbol{1};\boldsymbol{1} \rangle_{0,0}  |j;j-1;j-1; \boldsymbol{120};\boldsymbol{1}\rangle_{0,0} \oplus |j;j-1;j-1; \boldsymbol{1};\boldsymbol{120}\rangle_{0,0} \Big )  }
    \\
    &\oplus  \Big (|j+\tfrac12;j-\tfrac32;j-1\rangle_{0,0} \oplus |j+\tfrac12;j-\tfrac12;j-1 \rangle_{0,0} \oplus |j+\tfrac12;j-1;j-\tfrac32 \rangle_{0,0}  
      \\
      & \qquad \oplus |j +\tfrac12;j-1;j-\tfrac12\rangle_{0,0} \oplus |j-\tfrac12;j-\tfrac32;j-1\rangle_{0,0} \oplus |j-\tfrac12;j-\tfrac12;j-1 \rangle_{0,0} 
      \\
      &\qquad \oplus |j-\tfrac12;j-1;j-\tfrac32 \rangle_{0,0} \oplus |j -\tfrac12;j-1;j-\tfrac12\rangle_{0,0} \Big ) \otimes \overline{ \Big ( |j;j-1;j-1; \boldsymbol{128};\boldsymbol{1}\rangle_{0,0}}
      \\
      & \qquad\overline{\oplus |j;j-1;j-1; \boldsymbol{1};\boldsymbol{128}\rangle_{0,0} \oplus |j;j-1;j-1; \boldsymbol{16};\boldsymbol{16}\rangle_{0,0} \Big )  } \, ,
    \end{aligned}
\end{equation}
where here $I'=\{ j  \in \mathbb{Z}/2 \, , \, j\geq 1, \, j \leq \text{min}(\frac{k_s+1}{2},\frac{k_s^1}{2},\frac{k_s^2}{2})]\}$.
As in the previous case, we have a non-trivial contribution coming from the world-sheet bosons whose $q-$expansion gives
\begin{equation}
\begin{aligned} \label{eq:antiholobosS3S1}
     \bar \chi^{+}_{j,0} \  \bar \chi_{j-1} \ \bar \chi_{j-1} \ \bar \eta^{-1} \ \bar \eta^2 \sim & \Big ( \bar z+ \bar z^2 + \bar z^3 +\ldots + ( 1+   \bar z + 2 \bar z^2+\ldots ) \bar q+ O(\bar q^2) \Big ) 
     \\
     &\Big ( 1+ (\bar z_1^{-1} + 1+ \bar z_1) \bar q + O(\bar q^2)\Big ) \Big ( 1+ (\bar z_2^{-1} + 1+ \bar z_2) \bar q + O(\bar q^2)\Big )  \, ,
\end{aligned}
\end{equation}
where we have again omitted the overall contribution to the zero point energy. 
As before, the $O(\bar q^{0})$ term in the first line identifies the representation $D_j^+$, while the $O(\bar q)$ term the $D_{j-1}^+ \oplus D_{j+1}^+$ representation.
As before, for $j=1$ the reducible representation decomposes as $D_0^+=\mathbf{1} \oplus D_1^+$. In the second line we instead read the adjoint of $\mathfrak{su(2)} \oplus \mathfrak{su(2)}$. 
Therefore, for the $j=1$ spectrum, we have
\begin{equation}
\begin{aligned}
        &\Big (|0;0;0\rangle_{0,0} \oplus |2;0;0\rangle_{0,0} \oplus |1;1;0\rangle_{0,0} \oplus |1;0;1\rangle_{0,0} \oplus |1;0;0\rangle_{0,0}\Big)
        \\
        &\otimes \overline{\Big (|0;0;0;\boldsymbol{1};\boldsymbol{1}\rangle_{0,0} \oplus |2;0;0;\boldsymbol{1};\boldsymbol{1}\rangle_{0,0} \oplus |1;1;0;\boldsymbol{1};\boldsymbol{1}\rangle_{0,0} \oplus |1;0;1;\boldsymbol{1};\boldsymbol{1}\rangle_{0,0}}
        \\
        & \qquad \overline{\oplus |1;0;0;\boldsymbol{1};\boldsymbol{1}\rangle_{0,0} \oplus |1;0;0;\boldsymbol{120};\boldsymbol{1}\rangle_{0,0} \oplus |1;0;0;\boldsymbol{1};\boldsymbol{120}\rangle_{0,0} \Big)}\, .
        \end{aligned}
\end{equation}
The contribution from the R sector instead does not imply any further subtleties and reads
\begin{equation}
\begin{aligned}
        &|\tfrac32;\tfrac12;\tfrac12\rangle_{0,0}  \oplus    |\tfrac12;\tfrac12;\tfrac12\rangle_{0,0} \otimes \overline{ \Big ( |1;0;0; \boldsymbol{128};\boldsymbol{1}\rangle_{0,0}\oplus |1;0;0; \boldsymbol{1};\boldsymbol{128}\rangle_{0,0} \oplus |1;0;0; \boldsymbol{16};\boldsymbol{16}\rangle_{0,0} \Big )  } \, .
      \end{aligned}
\end{equation}
Therefore from these results we can conclude that there are no massless states arising from the R sector but we have the dilaton, the graviton, gauge fields transforming in the adjoint representation of $\mathfrak{su(2)} \oplus \mathfrak{su(2)} \oplus \mathfrak{so(16)} \oplus \mathfrak{so(16)}$,  and we have massless scalars coming from 
\begin{equation}\label{eq:Wilsonlines}
\begin{aligned}
        &\Big (|1;1;0\rangle_{0,0} \oplus |1;0;1\rangle_{0,0} \oplus |1;0;0\rangle_{0,0}\Big)
        \\
        &\otimes \overline{\Big ( |1;1;0;\boldsymbol{1};\boldsymbol{1}\rangle_{0,0} \oplus |1;0;1;\boldsymbol{1};\boldsymbol{1}\rangle_{0,0} \oplus |1;0;0;\boldsymbol{1};\boldsymbol{1}\rangle_{0,0}}
        \\
        & \qquad \overline{\oplus |1;0;0;\boldsymbol{120};\boldsymbol{1}\rangle_{0,0} \oplus |1;0;0;\boldsymbol{1};\boldsymbol{120}\rangle_{0,0} \Big)}\, .
        \end{aligned}
\end{equation}
These states correspond to the deformation moduli associated with the internal manifold and gauge degrees of freedom. This means that we can turn on a non-trivial VEV for the scalars along the Cartan generators that we can use to deform the theory. These correspond to the well-known Wilson line and we can show that there exist non-trivial directions which jeopardise the stability of the vacuum. This will be discussed in the next Section. 

\paragraph{Turning on a Wilson line} \label{ssssec:Wilson}

We can now move to the possibility of deforming the spectrum by turning on non-trivial Wilson lines. The analysis is performed for the AdS$_3 \times S^3 \times S^3 \times S^1$ background for technical simplicity but no conceptual obstruction prevents us from turning on Wilson lines for the AdS$_3 \times S^3 \times T^4$ background as well.

In \eqref{eq:Wilsonlines}, we can identify the subset of the massless scalars 
\begin{equation}
    |1;0;0\rangle_{0,0} \otimes \overline{\Big ( |1;0;0;\boldsymbol{120};\boldsymbol{1} \rangle_{0,0}  \oplus |1;0;0;\boldsymbol{1};\boldsymbol{120} \rangle_{0,0}  \Big)}
\end{equation}
as the Cartan directions of $\mathfrak{so(16)} \oplus \mathfrak{so(16)}$ algebra. These thus correspond to Wilson lines that we can use to deform the world-sheet theory. In particular, since the world-sheet RNS fermions and the right-moving bosons are not coupled to the $\mathfrak{sl}(\mathfrak{2},\mathbb{R})$  bosons, introducing these Wilson lines follows the same steps as the flat space \cite{Ginsparg:1986wr} case. The partition function for the lattice of signature $(1,17)$ thus becomes in the hamiltonian picture
\begin{equation}
    \Gamma_{(1,17)} \big [ \substack{i \\ j} \big]= \frac{1}{\eta \bar \eta^{17}}\sum_{m,n} \sum_{\lambda} q^{\frac{\alpha'}{4} \big ( \frac{m-\lambda \cdot A_9-A_9 \cdot A_9 n/2}{R}+\frac{n R}{\alpha'}\big )^2} \bar q^{\frac{\alpha'}{4} \big ( \frac{m-\lambda \cdot A_9-A_9 \cdot A_9 n /2}{R}-\frac{n R}{\alpha'}\big )^2} \bar q^{\frac12(\lambda+ A_9 n)^2} \, ,
\end{equation}
where $\lambda$ in our case is a vector belonging to the lattices $(i,j)+ D_{8}\oplus D_8$, with $i,j=o,v,s,c$ denoting the conjugacy class of the $\mathfrak{so(2n)}$ algebra. In the lagrangian picture, the lattice becomes 
\begin{equation}\label{eq:generallattice}
     \Gamma_{(1,17)} \big [ \substack{i_1 \\ i_2} \big]= \frac{R}{4 \alpha' \sqrt{\tau_2}\eta \bar \eta^{17}}\sum_{\tilde{m},n} e^{- \pi \frac{R^2}{\tau_2 \alpha'}|\tilde{m}+\tau n|^2} \prod_{a=1}^2 \bigg (\sum_{\sigma^a=0,1} e^{i \pi \sigma^a \alpha^a_{i_a}} \prod_{I=1}^8 \theta\big [\substack{\rho^a_{i_a}/2+A_9^I n \\ \sigma^a/2+A_9^I \tilde{m} } \big ] e^{-i \pi (\tilde{m}n {A^I_9}^2 +n\sigma^a A^I_9)} \bigg ) \, ,
\end{equation}
where $\rho_{i}$ and $\alpha_{i}$ are quantities depending on the conjugacy class of $\mathfrak{so(2n)}$, and in particular read $\rho_{i}=0$ if ${i}= (o),(v)$ or $\rho_{i}=1$ if ${i}=(s),(c)$ while $\alpha_{{i}}=0,1,0,1$ of $i=(o),(v),(s),(c)$ respectively. In particular for the $\text{Spin}(16)\times \text{Spin}(16)\rtimes \mathbb{Z}_2$ theory, the GSO projection couples  
\begin{equation}
    \begin{aligned}
        &V_8 \leftrightarrow (o,o)+(s,s) \, ,
        \\
        &O_8 \leftrightarrow (v,c)+(c,v) \, ,
        \\
        &S_8 \leftrightarrow (o,s)+(s,o) \, ,
        \\
        &C_8 \leftrightarrow (v,v)+(c,c) \, ,
    \end{aligned}
\end{equation}
where the $\mathfrak{so(8)}$ characters have to be decomposed in terms of the $\widehat{\mathfrak{so(2)}}_1 \otimes \widehat{\mathfrak{so(3)}}_1 \otimes \widehat{\mathfrak{so(3)}}_1$ characters. Therefore we have in general eight independent lattices that we have to analyse. 
To show that the theory develops tachyons when non-trivial Wilson lines are turned on, it is enough to choose an example on which we can perform analytic computations. For instance, we can introduce the following Wilson line \cite{Fraiman:2023cpa} at $R^2=\frac{\alpha'}{18}$
\begin{equation}
    A=(1,0^7;(\tfrac13)^8) \, , 
\end{equation}
along the Cartan directions of the $\mathfrak{so(16)} \oplus \mathfrak{so(16)}$ algebra. The expression thus becomes
\begin{equation}
\begin{aligned}
     \Gamma_{(1,17)} \big [ \substack{i_1 \\ i_2} \big]= \frac{1}{12\sqrt{2}   \sqrt{\tau_2}\eta \bar \eta^{17}}\sum_{\tilde{m},n} e^{- \pi \frac{1}{18\tau_2  }|\tilde{m}+\tau n|^2} &\bigg (\sum_{\sigma^1=0,1} e^{i \pi \sigma^1 (\alpha^1_{i_1}-n)} \theta\big [\substack{\rho^1_{i_1}/2 \\ \sigma^1/2+ \tilde{m} } \big ] \Big( \theta\big [\substack{\rho^1_{i_1}/2 \\ \sigma^1/2} \big ] \Big )^7 \bigg ) 
     \\
     &\bigg (\sum_{\sigma^2=0,1} e^{i \pi \sigma^2 (-n\frac83+\alpha^2_{i_2})}  \Big( \theta\big [\substack{\rho^2_{i_2}/2 +n/3\\ \sigma^2/2+\tilde{m}/3} \big ] \Big )^8 \bigg ) e^{-i \pi n \tilde{m} \frac89} \, .
     \end{aligned}
\end{equation}
To read the spectrum, we have to go back to the Hamiltonian picture by performing a Poisson resummation. However, in order for this to be done properly, we have to split $\tilde{m}$ and $n$ mod $18$ by $\tilde{m}=18 \tilde{r}+k$ and $n=18s+\ell$, with $k,\ell=0,1,2$, resulting into the final expression
\begin{equation}
\begin{aligned}
     \Gamma_{(1,17)} \big [ \substack{i_1 \\ i_2} \big]= &\frac{1}{72 \eta \bar \eta^{17}} \sum_{\sigma^1,\sigma^2} \sum_{r,s} \sum_{k,\ell=0}^{17} q^{\frac12 \big (\sqrt{18}s+\frac{\ell+r}{\sqrt{18}}\big)^2 }\bar q^{\frac12 \big (\sqrt{18}s+\frac{\ell-r}{\sqrt{18}}\big)^2 } e^{2 \pi i \frac{k}{18} (r + l) -i \pi \ell (\sigma^1 + \frac83 \sigma^2)}
     \\
     &\qquad e^{i \pi  (\alpha^1_{i_1}  \sigma^1 + \alpha^2_{i_2}  \sigma^2)}  \Big( \bar \theta\big [\substack{\rho^1_{i_1}/2 \\ \sigma^1/2} \big ] \Big )^8   \Big( \bar \theta\big [\substack{\rho^2_{i_2}/2 +\ell/3\\ \sigma^2/2+k/3} \big ] \Big )^8   \, .
     \end{aligned}
\end{equation}
We can now describe the contribution form the fermionic oscillators and the gauge degrees of freedom by analysing the contributions piece by piece. We can start from the piece paired to the vector given by $(i_1,i_2)=(o,o)$ implying $\rho=(0,0)$ and $\alpha=(0,0)$ and $(i_1,i_2)=(s,s)$ implying $\rho=(1,1)$ and $\alpha=(0,0)$. This contributes as 
\begin{equation}
\begin{aligned}
     V_8 \Big (\Gamma_{(1,17)} \big [ \substack{(o) \\ (o)} \big]+&\Gamma_{(1,17)} \big [ \substack{(s) \\ (s)} \big] \Big)
     \\
     &\sim  8 q^\frac12 + 16 q + 144 q^\frac12 \bar q^\frac12 + 288 q \bar q^\frac12 +  2176 q^\frac12 \bar q 
     \\
     &\qquad \qquad \qquad+ 4352  q \bar q + 7168 q^\frac12 q^\frac32 + 14336 q \bar q^\frac32 \ldots  \, ,
\end{aligned}
\end{equation}
where we have ignored the contribution to the zero point energy. The adjoint of the gauge group is related to the coefficient of $O(q^{\frac12} \bar q)$, namely $8\cdot 272$. Indeed, the full AdS$_3$ partition function (after decomposing the $V_8$ characters of $\widehat{\mathfrak{so(8)}}_1$ into $\widehat{\mathfrak{so(2)}}_1 \oplus \widehat{\mathfrak{so(3)}}_1 \oplus \widehat{\mathfrak{so(3)}}_1$ characters) gives rise to 
\begin{equation}
\begin{aligned}
        &\Big (|0;0;0\rangle   \oplus |2;0;0\rangle \oplus |1;1;0\rangle  \oplus |1;0;1\rangle  \oplus |1;0;0\rangle \Big)
        \\
        &\otimes \overline{\Big (|0;0;0;\boldsymbol{1};\boldsymbol{1}\rangle  \oplus |2;0;0;\boldsymbol{1};\boldsymbol{1}\rangle  \oplus |1;1;0;\boldsymbol{1};\boldsymbol{1}\rangle  \oplus |1;0;1;\boldsymbol{1};\boldsymbol{1}\rangle }
        \\
        & \qquad \overline{  \oplus |1;0;0;\boldsymbol{120};\boldsymbol{1}\rangle  \oplus |1;0;0;\boldsymbol{1};\boldsymbol{153}\rangle  \Big)}\, ,
        \end{aligned}
\end{equation}
where $\boldsymbol{273}=\boldsymbol{120}\oplus \boldsymbol{153}$ corresponds to the adjoint representation of $\mathfrak{so}(16)\oplus  \mathfrak{so}(18)$.

\noindent Similarly, we can proceed with the other terms. Looking at the lattices paired with $S_8$ we have $(i_1,i_2)=(o,s)$ implying $\rho=(0,1)$ and $\alpha=(0,0)$ and $(i_1,i_2)=(s,o)$ implying $\rho=(1,0)$ and $\alpha=(0,0)$ which gives 
\begin{equation}
\begin{aligned}
     S_8 \Big (\Gamma_{(1,17)} \big [ \substack{(o) \\ (s)} \big]+&\Gamma_{(1,17)} \big [ \substack{(s) \\ (o)} \big] \Big)
     \\
     &\sim   3072 q^\frac12 \bar q + 6144 q \bar q+8960  q^\frac12 q^\frac32 + 17920 q \bar q^\frac32 \ldots  \, ,
\end{aligned}
\end{equation}
where the Ramond ground state comes from the term with conformal weights $q^\frac12 \bar q$ given by $8\cdot (128+256)$. Putting everything in the partition function, we thus obtain
\begin{equation}
\begin{aligned}
        &|\tfrac32;\tfrac12;\tfrac12\rangle   \oplus    |\tfrac12;\tfrac12;\tfrac12\rangle  \otimes \overline{ \Big ( |1;0;0; \boldsymbol{128};\boldsymbol{1}\rangle \oplus |1;0;0; \boldsymbol{1};\boldsymbol{256}\rangle   \Big )  } \, .
      \end{aligned}
\end{equation}
This allows to read the spinorial representations of $\mathfrak{so}(16)\oplus \mathfrak{so}(18)$.

\noindent We can proceed with the analysis of the $C_8$ sector, which is paired with $(i_1,i_2)=(v,v)$ implying $\rho=(0,0)$ and $\alpha=(1,1)$ and $(i_1,i_2)=(c,c)$ implying $\rho=(1,1)$ and $\alpha=(1,1)$. This gives 
\begin{equation}
\begin{aligned}
     C_8 \Big (\Gamma_{(1,17)} \big [ \substack{(v) \\ (v)} \big]+&\Gamma_{(1,17)} \big [ \substack{(c) \\ (c)} \big] \Big)
     \\
     &\sim  1024 q^\frac12 \bar q + 2048 q \bar q+ 1024  q^\frac12 q^\frac32 + 2048 q \bar q^\frac32 \ldots  \, .
\end{aligned}
\end{equation}
Hence, the R ground state corresponding to the $O(q^\frac12 \bar q)$ term transforms into the $(\boldsymbol{128},\boldsymbol{1})$ representation of $\mathfrak{so}(16)\oplus\mathfrak{so}(18)$ as can be seen from
\begin{equation}
\begin{aligned}
        &|\tfrac32;\tfrac12;\tfrac12\rangle   \oplus    |\tfrac12;\tfrac12;\tfrac12\rangle  \otimes \overline{ \Big ( |1;0;0; \boldsymbol{128};\boldsymbol{1}\rangle  \Big )  } \, .
      \end{aligned}
\end{equation}

\noindent Finally, we have to analyse the $(i_1,i_2)=(v,c)$ sector, implying $\rho=(0,1)$ and $\alpha=(1,1)$ and the $(i_1,i_2)=(c,v)$ sector, implying $\rho=(1,0)$ and $\alpha=(1,1)$ . The contribution is 
\begin{equation}
\begin{aligned}
     O_8 \Big (\Gamma_{(1,17)} \big [ \substack{(v) \\ (c)} \big]+&\Gamma_{(1,17)} \big [ \substack{(c) \\ (v)} \big] \Big)
     \\
     &\sim  16 \bar q^\frac12 + 32 q^\frac12 \bar q^\frac12 + 576 q \bar q^\frac12+ 288 \bar q + 576 q^\frac12 \bar q \\
     & \qquad \qquad \qquad + 10368 q \bar q+256 \bar q^{\frac32} +512  q^\frac12 q^\frac32 + 9216 q \bar q^\frac32 \ldots  \, ,
\end{aligned}
\end{equation}
This allows us to read that there is a level-matched tachyon in the theory. Indeed, the NS ground state gives $\frac12-N-h=-\frac12$, hence $O(\bar q^{\frac12})$ in the holomorphic sector and $1-N-h=-\frac12$ in the anti-holomorphic sector, leading to the same expression for \eqref{eq:tachyonsS3S1} on both left-moving and right-moving sectors. Moreover, from the partition function we read that it transforms in the $(\boldsymbol{16},\boldsymbol{1})$ representation of $\mathfrak{so(16)}\oplus\mathfrak{so(18)}$. 
We have additional states arising from the term $O(q^{\frac12} \bar q)$ satisfying the mass-shell condition for the unflowed discrete representations. These latter states for $j=1$ thus correspond to 
\begin{equation}
    |1;0;0 \rangle \otimes \overline{|1;0;0;\boldsymbol{16};\boldsymbol{18}\rangle} \, ,
\end{equation}
identifying massless scalars in the $(\boldsymbol{16},\boldsymbol{18})$ of $\mathfrak{so(16)}\oplus\mathfrak{so(18)}$.
For a generic value of $j$ the low-energy spectrum thus reads
\begin{equation}
\begin{aligned}
    \bigoplus_{j \in I'}  &\Big (   |j-1;j-1;j-1\rangle  \oplus |j+1;j-1;j-1 \rangle  \oplus |j;j-2;j-1\rangle  \oplus |j;j-1;j-1 \rangle  
    \\
    &\qquad \oplus |j;j;j-1 \rangle  \oplus |j;j-1;j-2\rangle  \oplus |j;j-1;j-1 \rangle  \oplus |j;j-1;j \rangle  \Big )
    \\
    & \quad \otimes \overline{ \Big ( |j;j-1;j-1; \boldsymbol{120};\boldsymbol{1}\rangle  \oplus |j;j-1;j-1; \boldsymbol{1};\boldsymbol{153}\rangle  \Big )  }
    \\
    &\oplus \Big (   |j;j-1;j-1\rangle  \Big ) \otimes \overline{ \Big ( |j;j-1;j-1; \boldsymbol{16};\boldsymbol{18}\rangle  \Big )  }
    \\
    &\oplus  \Big (|j+\tfrac12;j-\tfrac32;j-1\rangle  \oplus |j+\tfrac12;j-\tfrac12;j-1 \rangle  \oplus |j+\tfrac12;j-1;j-\tfrac32 \rangle   
      \\
      & \qquad \oplus |j +\tfrac12;j-1;j-\tfrac12\rangle  \oplus |j-\tfrac12;j-\tfrac32;j-1\rangle  \oplus |j-\tfrac12;j-\tfrac12;j-1 \rangle  
      \\
      &\qquad \oplus |j-\tfrac12;j-1;j-\tfrac32 \rangle  \oplus |j -\tfrac12;j-1;j-\tfrac12\rangle  \Big ) \otimes \overline{ \Big ( |j;j-1;j-1; \boldsymbol{128};\boldsymbol{1}\rangle }
      \\
      & \qquad\overline{\oplus |j;j-1;j-1; \boldsymbol{1};\boldsymbol{256}\rangle  \oplus |j;j-1;j-1; \boldsymbol{128};\boldsymbol{1}\rangle  \Big )  } \, .
    \end{aligned}
\end{equation}
Notice that now we have an additional contributions coming from the states in \eqref{eq:O8charactersS3S1}, which are absent in the standard case without Wilson lines.

\begin{figure}[!htb]
   \begin{minipage}{0.48\textwidth}
     \centering
     \includegraphics[width=.7\linewidth]{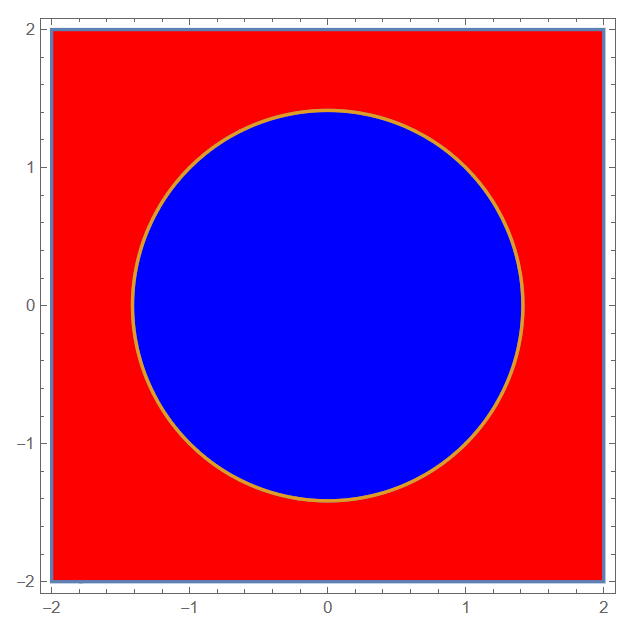}
     \caption{We show the plot of the variables $\{a_1,a_2\}$ for the choice of the radius $R^2=\alpha' (1-(a_1^2+a_2^2)/2)$ and Wilson line $A=(a_1,0^7;a_2,0^7)$ at $k_s^1=k_s^2=3$ and $j_1=j_2=0$. The blue region corresponds to tachyon-free points, while the red one to the tachyonic region.}\label{fig:moduli_space1}
   \end{minipage}\hfill
   \begin{minipage}{0.48\textwidth}
     \centering
     \includegraphics[width=.7\linewidth]{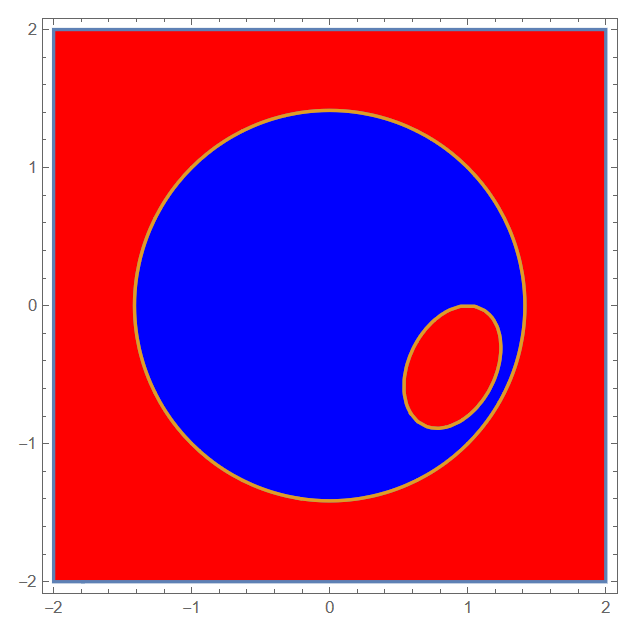}
     \caption{We show the plot of the variables $\{a_1,a_2\}$ for the choice of the radius $R^2=\alpha' (1-(a_1^2+a_2^2)/2)$ and Wilson line $A=(a_1,0^7;a_2,0^7)$ at $k_s^1=k_s^2=10^6$ and $j_1=j_2=0$. The blue region corresponds to tachyon-free points, while the red one to the tachyonic region.}\label{fig:moduli_space2}
   \end{minipage}
\end{figure}

This result explicitly shows the presence of tachyons for a specific choice of the Wilson line and the compactification radius, but they may appear also for different values of the moduli as they do in flat space \cite{Fraiman:2023cpa}. For instance, we can write down general formulas allowing us to identify tachyonic regions at tree-level. Indeed, tachyons are present if the NS ground state is level matched, which means that we need to impose
\begin{equation}
    h^L_{S^1}-\frac12=h^R_{S^1} -1 \, ,
\end{equation}
where we have used the superscript $L,R$ to distinguish among the contributions to the holomorphic and anti-holomorphic sectors
\begin{equation}
\begin{aligned}
    &h^L_{S^1}= \frac{\alpha'}{4} \Big ( \frac{m-\lambda \cdot A_9-A_9 \cdot A_9 n/2}{R}+\frac{n R}{\alpha'}\Big )^2 \, ,
    \\
    &h^R_{S^1}= \frac{\alpha'}{4} \Big ( \frac{m-\lambda \cdot A_9-A_9 \cdot A_9 n /2}{R}-\frac{n R}{\alpha'}\Big )^2 + \frac12(\lambda+ A_9 n)^2 \, .
    \end{aligned}
\end{equation}
The lattice vectors $\lambda \in (v,c)+(c,v)$, hence the minimal values for which they give rise to non-trivial expressions are $\lambda=(1,0^7;-\frac12,\frac12^7)$. Now we need to require that tachyons are allowed, which means asking  
\begin{equation}
    -1-4k_s j_1(1 + j_1)/k_s^1 -4k_s j_2(1 + j_2)/k_s^2 + 4k_s\Big (\frac12-h^{L}_{S^1}\Big ) \geq 0 \, .
\end{equation}
We can use these two conditions to investigate tachyonic regions of the moduli space. Generally speaking investigating the moduli space is hard, since at fixed AdS$_3$ curvature, it depends on nineteen parameters. Hence to proceed one can only choose suitable slices of the moduli space. Moreover, we notice that the AdS$_3$ curvature enters in these expressions in two ways: explictly and implicitly through the spins $j_1,j_2$ of the internal $\mathfrak{su(2)}$ since they are bounded from above by $(k_s^{1,2}-2)/2$. 
A simple numerical evaluation performed on the slice $A=(a_1,0^7;a_2,0^7)$ and $R^2=\alpha' (1-(a_1^2+a_2^2)/2)$ reported in fig. \ref{fig:moduli_space1} and \ref{fig:moduli_space2} reflects this. 
It would be interesting to make this analysis more systematic and to understand at which value of $k_s$ the tachyonic region is maximised.

The presence of tachyonic regions in the tree-level moduli space is already telling us that the theory has non perturbative instabilities. Indeed, even though there could be perturbatively stable points, nothing prevents tunnelling effects to unstable regions from taking place. A proper analysis though would require a precise evaluation of the tunnelling probability which is in general hard to put forward. Moreover, even though there are directions in the moduli space where tree-level tachyons do not appear, one-loop corrections may induce the presence of tachyons in the spectrum. To have a proper description of the structure of the one-loop corrected moduli space one should evaluate the $2$pt functions associated with these scalars on the torus and see whether the one-loop mass is below the BF bound. In \cite{Baykara:2022cwj,Fraiman:2023cpa}, considering the circle curvature to be much larger than the AdS$_3$ curvature, points developing negative mass scalars were found. However, they are still above the BF bound, thus ensuring perturbative stability. A precise world-sheet computation of the $2$pt amplitude would allow to see whethere these points are preserved even beyond the flat space approximation adopted in \cite{Baykara:2022cwj,Fraiman:2023cpa}. We leave this analysis to the possibly near future.

\section{Outlook}\label{sec:outlook}

In this paper, we have explored how non-supersymmetric strings are quantised on AdS$_3$ backgrounds. We have seen that, even though the $\text{Spin}(16) \times \text{Spin}(16) \rtimes \mathbb{Z}_2$ theory is free from tachyons, Wilson lines can be turned on, inducing the presence of such instabilities in the spectrum. This means that there are regions of the moduli space which are dangerous to which tunnelling effects may take place, making the vacuum non-perturbatively unstable. Nonetheless, as argued in \cite{Baykara:2022cwj,Fraiman:2023cpa} in the flat space limit, there could be points where the theory is perturbatively stable with directions developing negative mass scalars above the BF bound. A precise evaluation of the $2$pt ampltude on the torus would show if this picture persists also for other values of the AdS$_3$ curvature. 

In this analysis, we have considered only those theories that admit a tree-level AdS$_3$ background. However, in \cite{Baykara:2022cwj} an additional class of theories has been discussed, where the AdS$_3$ curvature contribution to the conformal anomaly is compensated by the one-loop tadpole. This, however, does not correspond to a CFT in the usual sense and lies beyond the available technology. Nonetheless, an explicit description of these theories would provide a deep insight into the nature of non-supersymmetric strings.

For the purpose of this paper, we have focused on the low-energy theory and on the presence of tachyons, which, as we have seen, only requires the unflowed sector to be studied. Indeed, the description of the flowed sector and of the massive string states was sketchy, and it will be interesting to tackle this issue in more detail. The massive spectrum is crucial if one wants to explore the possible holographic dual for these theories. In particular, it would be interesting to adapt the analyses performed for the bosonic string and the superstring known in the literature, both at a generic value of the level of the Ka$\check{\text{c}}$-Moody algebra \cite{Eberhardt:2019qcl, Eberhardt:2021vsx} and in the tensionless limit \cite{Gaberdiel:2018rqv,Eberhardt:2018ouy,Eberhardt:2019ywk,Gaberdiel:2024dva}. However, already the tensionless limit of the supersymmetric $ \text{E}_8 \times \text{E}_8 \rtimes \mathbb{Z}_2$ heterotic theory seems non-trivial, and a better understanding of this case would be necessary to gain a better understanding of the non-supersymmetric set-up as well. 

Moreover, we have seen that these spaces still have instabilities arising from the presence of Wilson lines. A possible direction to solve this issue would be applying the lesson that we have learnt in flat space by engineering non-geometric constructions \cite{Narain:1986qm,Baykara:2023plc,Baykara:2024tjr,Angelantonj:2024jtu}. In this case, if we have found a non-tachyonic point, no tunnelling effects can take place since other regions of the moduli space are forbidden. Alternatively, it would be interesting to study the euclidean continuation of AdS$_3$ discussed recently in \cite{Eberhardt:2025sbi}, where the spectrum was shown to be composed of $w\geq 1$ continuous representations. This means that no tachyons can be present and indeed already the bosonic string does not have such instabilities \cite{Eberhardt:2025sbi}. 

\section*{Acknowledgments}

The author is grateful to Ivano Basile, Marios Petropoulos and Augusto Sagnotti for useful discussions at various stages of the project, and to Carlo Angelantonj and Matthias Gaberdiel for useful discussions and comments on the manuscript. This work is funded by the European Union - NextGenerationEU/PNRR mission 4.1; CUP: C93C24004950006.  

\appendix

\section{Theta functions and characters} \label{app:thetaandcharacters}

In the Appendix, we report the properties of characters and elliptic functions involved. 
In particular, as described in Section \ref{sec:sl2Rgeneralities}, one needs the expression of the complete refined characters to regularise the one-loop contribution to the path integral, still preserving their modular properties \cite{Kac:1984mq,Kato:2000tb}. A straightforward computation allows to identify the characters associated with the spectrally flowed lowest weight discrete representation
\begin{equation}
\begin{aligned} \label{eq:flowedcharactersdiscrete}
    \chi_{j,w}^+(u,z,\tau) &= \text{tr} \ q^{ L_0 - w J_0^3-  \frac{k \, w^2}{4} - \frac{3 k}{24(k-2)} } \ z^{J_0^3+\frac{w \, k}{2} } \ e^{2\pi i u t}
    \\
    &= e^{2\pi i u k} (-1)^{w} \frac{ q^{-\frac{1}{k-2} \big (j-\frac12 +\frac{k-2}{2}w \big )^2} \ e^{2 \pi i y \big (j-\frac12 +\frac{k-2}{2}w \big ) }   }{i \, \Jtheta{1/2}{1/2}{-y}{\tau}}\, ,
\end{aligned}
\end{equation}
where we have used the definitions of the chemical potential of the Cartan generator $z=e^{2 \pi i y}$ and of central element of the Ka$\check{\text{c}}$-Moody algebra $t$ admitting $k$ as eigenvalue. In the expressions, the flavoured oscillators give rise to the Jacobi theta function with $1/2$ on both upper and lower characteristics following from the definition 
\begin{equation}
    \Jtheta{a}{b}{y}{\tau}= \sum_{n \in \mathbb{Z}} q^{\frac12 (n+a)^2} e^{2 \pi i (n+a)(y+b)}
\end{equation}
that by means of the Jacobi triple product identity becomes
\begin{equation}
\begin{aligned}
     \Jtheta{a}{b}{y}{\tau}&= e^{2 \pi i a( y+b)} q^{\frac{a^2}{2}} \prod_{n=1}^{\infty} \big ( 1- q^n \big ) \big ( 1+ e^{2 \pi i (y+b)} q^{n+a-\frac12} \big ) \big ( \big ( 1+ e^{-2 \pi i (y+b)} q^{n-a-\frac12} \big )  \, .
\end{aligned}
\end{equation}
A general modular transformation described by the element
\begin{equation}
    \begin{pmatrix}
        m & n
        \\
        p & s
    \end{pmatrix} \, , \quad \in \quad \text{PSL}(2,\mathbb{Z}) \, ,
\end{equation}
acts on the triple $(\tau, y,u) $ as follows 
\begin{equation}
    (\tau, y, u) \to \bigg ( \frac{m \tau + n}{p \tau + s}, \frac{y}{p \tau +s}, u + \frac{p y^2}{4(p\tau +s)}\bigg ) \, .
\end{equation}
Such transformations, combined with the well-known modular transformations of the theta functions 
\begin{equation}
    \begin{aligned}
        & \Jtheta{a}{b}{\frac{y}{\tau}}{-\frac{1}{\tau}}= \sqrt{- i \tau} \, e^{2 \pi i a b}  e^{\frac{i \pi y^2}{\tau}}\Jtheta{b}{-a}{y}{\tau} \, ,
        \\
        &\Jtheta{a}{b}{y}{\tau+1}=e^{- i \pi a(a-1)} \Jtheta{a}{a+b-1/2}{y}{\tau} \, ,
    \end{aligned}
\end{equation}
allow us to verify the modular invariance of the discrete sector of the partition function described in eq. \eqref{eq:bosAdS3} explicitly. 
The continuous representations are captured by the characters 
\begin{equation}\label{eq:flowedcharacterscontinuous}
\begin{aligned}
    \chi_{j,w}^{\alpha}(u,z,\tau) &=\text{tr} \ q^{ L_0 - w J_0^3-  \frac{k \, w^2}{4} - \frac{3 k}{24(k-2)} } \ z^{J_0^3+\frac{w \, k}{2} } \ e^{2\pi i u t} 
    \\
    &= \frac{q^{-\frac{(j-1/2)^2}{k-2}+ \frac{k \, w^2}{4}}}{\eta(\tau)^3} \, 2 \pi e^{2\pi i u k} \sum_{\ell} e^{-2 \pi i \ell \big (\alpha+\frac{k w}{2} \big)} \delta(\ell'+y-\tau w)    \, ,
\end{aligned}
\end{equation}
where we have used the Dedekind $\eta$ function
\begin{equation}
\eta(\tau)= q^{\frac{1}{24}} \prod_{n=1}^{\infty} ( 1- q^n) \, .
\end{equation}
Taking advantage of the properties of the $\delta$-function and of the transformation properties under the $S$ and $T$ generators of Dedekind eta function  
\begin{equation}
    \begin{aligned}
        & \eta(-\frac{1}{\tau})= \sqrt{-i \tau} \eta(\tau) \, ,
        \\
        & \eta(\tau + 1)= e^{\frac{i \pi}{12}} \eta(\tau) \, ,
    \end{aligned}
\end{equation}
one can verify the modular invariance of the partition function in eq. \eqref{eq:bosAdS3} for the continuous sector.

To discuss the situation for the world-sheet fermions on AdS$_3$ few comments are in order. In the NS sector, the zero mode algebra is not affected \cite{Ferreira:2017pgt}, and hence the highest weight states of the Virasoro algebra correspond to \begin{equation}
    J^a_n |j,m\rangle=0 \, , \quad n>0 \, , \qquad \psi^a_r|j,m\rangle=0 \, , \quad r\geq \tfrac12 \, ,
\end{equation}
both for the discrete and continuous representations.
In the R sector, however, the presence of the zero modes multiplies the representations of the ground states with the spinorial representation labelled by $s_0=\pm$. The action of the shifted current becomes \cite{Ferreira:2017pgt}
\begin{equation}
   \Big ( \mathcal{J}^3_0+\frac{1}{k}:\psi^+ \psi^-:_0 \Big ) |j,m,s_0 \rangle= J^3_0|j,m,s_0 \rangle=\big (m+\frac{s_0}{2} \big )|j,m,s_0 \rangle \, ,
\end{equation}
where 
\begin{equation}
    :\psi^+ \psi^-:_0 |s_0\rangle=\frac12 \big [ \psi_0^+, \psi_0^- \big ]|s_0\rangle= k \frac{\sigma^3}{2} |s_0\rangle= k \frac{s_0}{2} |s_0\rangle \, .
\end{equation}
To obtain the contribution of fermionic states with a given spin structure, we need to specify the effect of the spectral flow. We can compute this piece by taking the trace of the flowed operators over the unflowed Hibert space. This means that the fermion number needed to specify the periodic boundary conditions along one of the two cycles of the torus is spectrally flowed as well. A straightforward computation when the proper normal ordering is taken into account shows \cite{Giribet:2007wp,Giribet:2018ada} 
\begin{equation}
\begin{aligned}
    \tilde{F}&= \frac{1}{k}\sum_{r>0}  \big ( \tilde{\psi}^+_{-r}\tilde{\psi}^-_{r} + \tilde{\psi}^-_{-r}\tilde{\psi}^+_{r} - 2 \tilde{\psi}^3_{-r}\tilde{\psi}^3_{r} \big )
    \\
    &=F+w \, .
    \end{aligned}
\end{equation}
This gives rise to the GSO projection, which is therefore affected according to whether the spectral flow parameter is even or odd.
We can now compute the contribution of the fermions with the superghosts compensating the Cartan direction with a given choice of the spin structure. Taking anti-periodic boundary conditions on the "time"-like cycle we have  as 
\begin{equation}
    \begin{aligned}\label{eq:spinstructuresNS}
    \mathcal{Z}_{\psi} \big [ \substack{1/2 \\ \beta} \big ] &= \text{tr}_{NS} \ e^{2\pi i\beta (F+ w)} q^{L_0-w J^3_0 -\frac{k w^2}{4}-\frac{c}{24}} z^{J^3_0+ \frac{k w}{2}} 
    \\
        &= \prod_{r=\frac12}^{\infty} \prod_{a=\pm} \sum_{\lambda_{a,r}=0}^1 \langle \lambda_{a,r};0 | (-1)^{\beta(w+\lambda_{a,r})} q^{r \lambda_{a,r}-\frac{kw^2}{4}-w \lambda_{a,r}-\frac{3}{48}} z^{\rho_a \lambda_{a,r}+\frac{kw}{2}} |0;\lambda_{a,r} \rangle
        \\
        &= (-1)^{\beta w} q^{-\frac{kw^2}{4}-\frac{3}{48}} z^{\frac{kw}{2}} \prod_{r=\frac12}^{\infty} \prod_{a=\pm} (1+q^{r-w \rho^a} z^{\rho_a}e^{2\pi i \beta}) 
        \\
        &= i^{\frac32} q^{-\frac{(k+2)w^2}{4}} z^{\frac{(k+2)w}{2}} \frac{\Jtheta{1/2}{\beta}{y}{\tau}}{\eta} \, ,
\end{aligned}
\end{equation}
where $\lambda_{a,r}$ denotes the number of the oscillators $\psi^a_{-r}$ acting on the NS ground state with $a=\pm$ and $\rho^{\pm}=\pm$ is the eigenvalue with respect to $J^3_0$. 
Similarly, for periodic boundary conditions we have
\begin{equation}
    \begin{aligned}\label{eq:spinstructuresR}
    &\mathcal{Z}_{\psi} \big [ \substack{0 \\ \beta} \big ] = \text{tr}_{R} \ e^{2\pi i\beta (F+ w)} q^{L_0-w J^3_0 -\frac{k w^2}{4}-\frac{c}{24}} z^{J^3_0+ \frac{k w}{2}} 
    \\
       &= \prod_{r=1}^{\infty} \prod_{a=\pm} \sum_{\lambda_{a,r}=0}^1 \sum_{s_0=\pm 1}\langle \lambda_{a,r} ,s_0 | (\sigma^3)^{2\beta}(-1)^{\beta(w+\lambda_{a,r})} q^{r \lambda_{a,r}-\frac{kw^2}{4}-w \lambda_{a,r}-\frac{3}{48}+\frac{3}{16}+\frac{w s_0}{2}} z^{\rho_a \lambda_{a,r}+\frac{kw}{2}-\frac{s_0}{2}} |\lambda_{a,r}, s_0 \rangle
        \\
        &=  (q^{\frac{w }{2}} z^{-\frac{1}{2}} +e^{2\pi i \beta} q^{-\frac{w }{2}} z^{\frac{ 1}{2}}) q^{-\frac{kw^2}{4}+\frac{1}{8}} z^{\frac{kw}{2}} \prod_{a=\pm}  \prod_{r=1}^{\infty} (1+q^{r-w \rho^a} z^{\rho_a}e^{2\pi i \beta}) 
        \\
        &= i^{\frac32} q^{-\frac{(k+2)w^2}{4}} z^{\frac{(k+2)w}{2}}  \frac{\Jtheta{0}{\beta}{y}{\tau}}{\eta} \, ,
\end{aligned}
\end{equation}
where we have taken into account the conformal weight of the $R$ ground state and the same notation for $\lambda_{a,r}$ and $\rho^a$. 
The powers of $q$ and $z$ in front of the $\theta$ function are shared with the bosonic sector of the Hilbert space and hence they all conspire to
\begin{equation}
    q^{-\frac{1}{k} \big (j-\frac12 +\frac{k}{2}w \big )^2} \ e^{2 \pi i y \big (j-\frac12 +\frac{k}{2}w \big ) } \, ,
\end{equation}
which is consistent with the result in \eqref{eq:bosAdS3} up to $k \to k+2$. Hence, using the modular properties of the $\theta$ functions and introducing $u$, one can show the modular invariance of the partition function of the type 0B superstring and heterotic strings discussed in \ref{sec:type0} and \ref{sec:16x16}, as well as the type IIB superstring in \cite{Israel:2003ry}.

To read the spectrum is nonetheless convenient to express everything in terms of the characters. From eqs. \eqref{eq:spinstructuresNS} and \eqref{eq:spinstructuresR}, combined with the contribution of the remaining six fermions, we can write down the partition function in terms of the $\widehat{D}_{n,1}$ characters
\begin{equation}
    \begin{aligned}
        & O_{2n}(y|\tau)= \frac{\Jtheta{0}{0}{y}{\tau}^n+\Jtheta{0}{1/2}{y}{\tau}^n}{2 \eta(\tau)^n} \, ,
        \\
        &V_{2n}(y|\tau)= \frac{\Jtheta{0}{0}{y}{\tau}^n-\Jtheta{0}{1/2}{y}{\tau}^n}{2 \eta(\tau)^n} \, ,
        \\
        &S_{2n}(y|\tau)= \frac{\Jtheta{1/2}{0}{y}{\tau}^n+i^{-n}\Jtheta{1/2}{1/2}{y}{\tau}^n}{2 \eta(\tau)^n} \, ,
        \\
        &C_{2n}(y|\tau)= \frac{\Jtheta{1/2}{0}{y}{\tau}^n-i^{-n}\Jtheta{1/2}{1/2}{y}{\tau}^n}{2 \eta(\tau)^n} \, ,
    \end{aligned}
\end{equation}
and for $\widehat{B}_{n,1}$ characters
\begin{equation}
    \begin{aligned}
        & O_{2n+1}(y|\tau)= \frac{\Jtheta{0}{0}{y}{\tau}^{n+\frac12}+\Jtheta{0}{1/2}{y}{\tau}^{n+\frac12}}{2 \eta(\tau)^{n+\frac12}} \, ,
        \\
        &V_{2n+1}(y|\tau)= \frac{\Jtheta{0}{0}{y}{\tau}^{n+\frac12}-\Jtheta{0}{1/2}{y}{\tau}^{n+\frac12}}{2 \eta(\tau)^{n+\frac12}} \, ,
        \\
        &S_{2n+1}(y|\tau)= \frac{\Jtheta{1/2}{0}{y}{\tau}^{n+\frac12}}{\sqrt{2} \eta(\tau)^{n+\frac12}} \, .
    \end{aligned}
\end{equation}
The modular properties of the characters are well-known. Under the $S$ and $T$ generators of the modular group, the $\widehat{D}_{n,1}$ characters transform as
\begin{equation}
    S=\frac12\begin{pmatrix}
        1 & 1 & 1 & 1
        \\
        1 & 1 & -1 & -1
        \\
        1 & -1 & i^{-n} & -i^{-n}
        \\
        1 & -1 & -i^{-n} & i^{-n}
    \end{pmatrix} \, , \qquad T=e^{-i \pi n/12 } diag(1,-1,e^{i n \pi/4 },e^{i n \pi/4 }) \, ,
\end{equation}
while the $\widehat{B}_{n,1}$ characters transform as
\begin{equation}
    S=\frac12\begin{pmatrix}
        1 & 1 & \sqrt{2}
        \\
        1 & 1 & -\sqrt{2}
        \\
        \sqrt{2} & -\sqrt{2} & 0
    \end{pmatrix} \, , \qquad T=e^{-i \pi (n/12+1/24) } diag(1,-1,e^{i \pi (n+1/2) /4 }) \, .
\end{equation}
Finally the $S^3$ pieces contribute in terms of $\widehat{\mathfrak{su (2)}}_{k-2}$ with three free fermions. The characters describing the three free fermions have been already discussed. We only need to describe the world-sheet bosons whose expression is well-known and for $\widehat{A}_{n,k}$ reads 
\begin{equation}\label{eq:su2characters}
    \chi_j(u,y,\tau)= e^{-2 \pi i u k} \frac{\Theta_{2 j+1}^{(k+2)}(y|\tau)-\Theta_{-2 j-1}^{(k+2)}(y|\tau)}{\Theta_{1}^{(2)}(y|\tau)-\Theta_{-1}^{(2)}(y|\tau)} \, , \qquad j=0,\ldots, \frac{k}{2} \, ,
\end{equation}
where we have used the definition of the {\em generalised} theta functions
\begin{equation}
    \Theta^{(k)}_m(y|\tau)=\sum_{n \in \mathbb{Z}}q^{k (n+\frac{m}{2k})^2} e^{2 \pi i k y(n+\frac{m}{2k})} \, .
\end{equation}
The modular properties of the characters are inherited from the behaviour of the $\Theta$-function
\begin{equation}
    \begin{aligned}
        &\Theta_m^{(k)}(\frac{y}{\tau}|-\frac{1}{\tau})= \sqrt{\frac{-i \tau}{2k}} e^{2 \pi i k \frac{y^2}{4 \tau}} \sum_{m'=-k+1}^k e^{- 2 \pi i \frac{m m'}{2k}} \Theta^{(k)}_{m'}(y|\tau) \, ,
        \\
        &\Theta_m^{(k)}(y|\tau+1)= e^{ i \pi \frac{m^2}{2k^2} }\Theta_m^{(k)}(y|\tau) \, ,
    \end{aligned}
\end{equation}
and read
\begin{equation}
    S_{j j'}= \sqrt{\frac{2}{k+2}} \sin \bigg (\pi \frac{(2j+1) (2j'+1)}{k+2} \bigg)\, , \qquad T_{j j'}=e^{2\pi i \big (\frac{j(j+1)}{k+2}-\frac{3k}{24(k+2)} \big )} \delta_{jj'}\, .
\end{equation}
Although $\mathfrak{su(2)}$ and $\mathfrak{sl}(\mathfrak{2},\mathbb{R)}$ have the same complexification the expression of their characters is very different. This indeed due to the different Killing forms required to describe the two situations. Indeed in the $\mathfrak{su(2)}$ case, the presence of additional null states determines towers of states to be properly added and subtracted, leading to the expression of the characters \eqref{eq:su2characters}. For the $\mathfrak{sl}(\mathfrak{2},\mathbb{R})$ case the associated Killing form implies that no null states are present for the values which are compatible with he Maldacena-Ooguri bound \cite{Bakas:1991fs}, hence leading to the expressions in \eqref{eq:flowedcharactersdiscrete} and \eqref{eq:flowedcharacterscontinuous}. 

We can discuss the contribution coming from the $T^p$ term of the partition functions. The associated  fermions are all free fields so that they simply form the characters of $\widehat{\mathfrak{so(p)}}_1$. Hence, the only term that we have to discuss comes from the compact bosons which gives
\begin{equation}
    \Gamma_{(p,p)} (G,B)= \frac{1}{\big ( \sqrt{\tau_2} \eta \bar \eta)^{p}} \sum_{p_L, p_R \in \Lambda_{(p,p)}} q^{\frac12 {p_L}_a G^{ab} {p_L}_b } \bar q^{\frac12 {p_R}_a G^{ab} {p_R}_b} \, ,
\end{equation}
where $G,B$ correspond to the metric and the Kalb-Ramond field associated with the lattice. The definition of left and right momenta reads
\begin{equation}
    \begin{aligned}
        &{p_L}_a= \sqrt{\frac{\alpha'}{2}}  \Big ( m_a + \frac{1}{\alpha'} (G_{ab}-B_{ab})n^b \Big) \, ,
        \\
        &{p_L}_a =\sqrt{\frac{\alpha'}{2}}  \Big ( m_a - \frac{1}{\alpha'} (G_{ab}+B_{ab})n^b \Big)  \, .
    \end{aligned}
\end{equation}
Using Poisson summation, one can show that this contribution is modular invariant  under the $S$ and $T$ transformations.

\section{The supersymmetric \texorpdfstring{$\text{E}_8 \times \text{E}_8 \rtimes \mathbb{Z}_2$}{E8 x E8} heterotic string} \label{app:SUSYE8}

In this Appendix we discuss the supersymmetric heterotic theory with gauge group $E_8 \times E_8 \rtimes \mathbb{Z}_2$. As for the non-supersymmetric set-up, we study the AdS$_3 \times S^3 \times T^4$ and the AdS$_3 \times S^3 \times S^1$ backgrounds and we focus on the low-energy theory. In the first case we are expecting to realise the {\em small} $\mathcal{N}=(4,0)$) supersymmetry algebra $\mathfrak{psu}(1,1|2)$ and in the latter the {\em large} $\mathcal{N}=(4,0)$ supersymmetry algebra $D(2,1|\alpha)$.

\subsection{AdS\texorpdfstring{$_3 \times S^3 \times T^4$}{3 x S3 x T4}}

We can write he partition function by specifying the GSO projection characterising such a theory
\begin{equation}
\begin{aligned}
  \mathcal{T}_{E_8 \times E_8}(T^4) =  \int d \mu B_H(\tau, \bar \tau \, | \, T^4) F_{E_8\times E_8}  (\tau, \bar \tau \, | \, T^4) \, ,
          \end{aligned}
\end{equation}
where again $B_H(\tau, \bar \tau \, | \, T^4)$ is the contribution coming from the world-sheet bosons while  
\begin{equation}
\begin{aligned} 
  F_{E_8 \times E_8}  (\tau, \bar \tau \, | \, T^4)=& e^{-4 \pi i u}\Big ( V_2 O_2 O_4 + V_2 V_2 V_4 + O_2 V_2 O_4+ O_2 O_2 V_4 
        \\
        &  - S_2 S_2 S_4 - S_2 C_2 C_4 - C_2 S_2 C_4 - C_2 C_2 S_4 \Big )  \bar \chi_0 \bar \chi_0 \, ,
        \end{aligned}
\end{equation}
where we have denoted $\chi_0$ the only character of the $\widehat{\mathfrak{e}_{8}}_1$ algebra
\begin{equation}
    \chi_0= \frac{1}{2 \eta^8} \bigg ( \Jtheta{0}{0}{0}{\tau}^8 + \Jtheta{0}{1/2}{0}{\tau}^8 + \Jtheta{1/2}{0}{0}{\tau}^8 \bigg ) \sim q^{-\frac13} ( 1+ 248 q + \ldots ) \, .
\end{equation}
The low energy spectrum thus simply follows from the considerations made in the previous Section. Indeed, the left-moving sector corresponds precisely to that of the type IIB described in Section \ref{sec:type0}, while the right-moving one we have
\begin{equation}
    q^{-1} + (z^{-1} +z) + (z'^{-1} + z')+ 4 +248 +248 +\ldots
\end{equation}
from which we see that only the $q^0$ states are level-matched. The low-energy spectrum the reads
\begin{equation}
\begin{aligned}
    &\bigoplus_{j \in I}\Big ( (j-1)_{\text{short}} \oplus (j)_{\text{short}} \oplus 2(j-\tfrac12)_{\text{short}} \Big ) 
    \\
    & \qquad \qquad \quad \otimes \overline{\Big ( |j-1;j-1;0,0; \boldsymbol{1};\boldsymbol{1}\rangle_{0,0} \oplus |j+1;j-1;0,0; \boldsymbol{1};\boldsymbol{1}\rangle_{0,0} }
    \\
    &\qquad \qquad \qquad \overline{\oplus |j;j;0,0; \boldsymbol{1};\boldsymbol{1}\rangle_{0,0} \oplus |j;j-2;0,0; \boldsymbol{1};\boldsymbol{1}\rangle_{0,0}\oplus |j;j-1;\tfrac12,\tfrac12; \boldsymbol{1};\boldsymbol{1}\rangle_{0,0}}
    \\
    &\qquad \qquad \qquad\overline{\oplus|j;j-1;0,0; \boldsymbol{248};\boldsymbol{1}\rangle_{0,0} \oplus |j;j-1;0,0; \boldsymbol{1};\boldsymbol{248}\rangle_{0,0} \Big )  } \, ,
    \end{aligned}
\end{equation}
where the short representation for the algebra corresponds to
\begin{equation}
    (j)_{\text{short}}= |j;j\rangle \oplus 2 |j+\tfrac12;j-\tfrac12\rangle \oplus |j+1;j-1\rangle \, .
\end{equation}
For $j=1$, taking into account that the $D_0^+$ representation is reducible we have
\begin{equation}
\begin{aligned} 
       &\Big ( (0,0)_{\text{short}} \oplus 2 \cdot (\tfrac12,\tfrac12)_{\text{short}}  \oplus (1,1)_{\text{short}}  \Big ) \otimes 
       \overline{\Big ( |0; 0;0,0;\boldsymbol{1};\boldsymbol{1} \rangle  \oplus |2; 0;0,0 ;\boldsymbol{1};\boldsymbol{1}\rangle_{0,0}  \oplus |1;1;0,0; \boldsymbol{1} ;\boldsymbol{1} \rangle_{0,0}  }
       \\
       & \Big ( (0,0)_{\text{short}} \oplus 2 \cdot (\tfrac12,\tfrac12)_{\text{short}}  \oplus (1,1)_{\text{short}}  \Big ) \otimes 
       \overline{\Big ( |1;0;\tfrac12,\tfrac12; \boldsymbol{1}; \boldsymbol{1} \rangle_{0,0}  \Big ) } 
       \\
       &\Big ( (0,0)_{\text{short}} \oplus 2 \cdot (\tfrac12,\tfrac12)_{\text{short}}  \oplus (1,1)_{\text{short}}  \Big ) \otimes 
       \overline{\Big ( |1;0;0,0; \boldsymbol{248}; \boldsymbol{1} \rangle  \oplus |1;0;0,0;\boldsymbol{1};\boldsymbol{248} \rangle \Big )}
\end{aligned}
\end{equation}
where we recall
\begin{equation}
    \begin{aligned}
        &(0,0)_{\text{short}} =|0;0\rangle \, ,
        \\
        &(\tfrac12,\tfrac12)_{\text{short}} = |\tfrac12; \tfrac12 \rangle \oplus 2 |1;0\rangle \, ,
        \\
        &(1,1)_{\text{short}} = |1;1\rangle \oplus  2 |\tfrac12; \tfrac12 \rangle \oplus  |2;0\rangle \, ,
    \end{aligned}
\end{equation}
As expected, the theory is supersymmetric and hence it does not have any contribution from the unflowed continuous representations, which are responsible for the appearance of tachyons.

\subsection{AdS\texorpdfstring{$_3 \times S^3 \times S^3 \times S^1$}{3 x S3 x S3 x S1}}

\noindent A similar discussion can be performed for this background. The partition function corresponds to 
\begin{equation}
\begin{aligned}
  \mathcal{T}_{E_8 \times E_8}(S^3 \times S^1) =  \int d \mu \, B_H(\tau, \bar \tau \, | \, S^3 \times S^1) F_{E_8\times E_8}  (\tau, \bar \tau \, | \, S^3 \times S^1) \, ,
          \end{aligned}
\end{equation}
with $B_H(\tau, \bar \tau \, | \, S^3 \times S^1)$ the world-sheet bosons contribution, while  
\begin{equation}
\begin{aligned} 
  F_{E_8 \times E_8}  (\tau, \bar \tau \, | \, S^3 \times S^1)=&e^{-4 \pi i u} \Big (O_2  O_3  V_3 + V_2  V_3  V_3  + V_2  O_3  O_3   + O_2  V_3  O_3  
        \\
        &  -  S_2  S_3  S_3  -  C_2  S_3  S_3 \Big )  \bar \chi_0 \bar \chi_0 \, .
        \end{aligned}
\end{equation}
The contribution from the holomorphic sector corresponds to the one of the type IIB superstring, while the anti-holomorphic one comes from 
\begin{equation}
    q^{-1} + (z^{-1} +z) + (z_1^{-1} +1 + z_1)+ (z_2^{-1} +1 + z_2) +248 +248 +\ldots
\end{equation}
This allows us to read the low-energy spectrum as
\begin{equation}
\begin{aligned}
    &\bigoplus_{j \in I'} \Big ( (j-1;j-1;j-1)_{s} \oplus (j-\tfrac12;j-\tfrac12;j-\tfrac12)_{s} \Big ) 
    \\
    &\qquad \otimes \overline{ \Big ( |j-1;j-1;j-1; \boldsymbol{1};\boldsymbol{1}\rangle_{0,0} \oplus |j+1;j-1;j-1; \boldsymbol{1};\boldsymbol{1} \rangle_{0,0} \oplus |j;j-2;j-1; \boldsymbol{1};\boldsymbol{1}\rangle_{0,0}}
    \\
    &\qquad \quad \overline{ \oplus |j;j-1;j-1 ; \boldsymbol{1};\boldsymbol{1} \rangle_{0,0}  |j;j;j-1 ; \boldsymbol{1};\boldsymbol{1}\rangle_{0,0} \oplus |j;j-1;j-2; \boldsymbol{1};\boldsymbol{1}\rangle_{0,0} \oplus |j;j-1;j-1; \boldsymbol{1};\boldsymbol{1} \rangle_{0,0}}
    \\
    & \qquad  \quad \overline{\oplus |j;j-1;j; \boldsymbol{1};\boldsymbol{1} \rangle_{0,0}  |j;j-1;j-1; \boldsymbol{120};\boldsymbol{1}\rangle_{0,0} \oplus |j;j-1;j-1; \boldsymbol{1};\boldsymbol{120}\rangle_{0,0} \Big )  }\, ,
    \end{aligned}
\end{equation}
where  the representation of the {\em large} $\mathcal{N}=(4,0)$ superalgebra $D(2,1|\alpha)$ correspond to
\begin{equation}
\begin{aligned}
    (j;j_1;j_2)_s=&|j;j_1;j_2\rangle \oplus | j+\tfrac12;j_1+\tfrac12;j_2-\tfrac12\rangle \oplus |j+\tfrac12;j_1-\tfrac12;j_2+\tfrac12\rangle \oplus |j+\tfrac12;j_1-\tfrac12;j_2-\tfrac12\rangle 
    \\
    &\oplus |j+1;j_1-1;j_2\rangle \oplus |j+1;j_1;j_2\rangle \oplus |j +1;j_1;j_2-1\rangle\oplus |j+\tfrac32;j_1-\tfrac12;j_2-\tfrac12\rangle \, .
    \end{aligned}
\end{equation}
At $j=1$, and hence $j_1=j_2=0$, the spectrum reads
\begin{equation}
\begin{aligned} 
       &\Big ( (0,0,0)_s \oplus (\tfrac12,\tfrac12,\tfrac12)_s \Big ) \otimes 
       \overline{\Big ( |0; 0;0;\boldsymbol{1};\boldsymbol{1} \rangle  \oplus |2; 0;0 ;\boldsymbol{1};\boldsymbol{1}\rangle_{0,0}  \oplus |1;1;0; \boldsymbol{1} ;\boldsymbol{1} \rangle_{0,0} }
       \\
       & \Big ( (0,0,0)_s \oplus (\tfrac12,\tfrac12,\tfrac12)_s \Big ) \otimes \overline{\Big (  |1;0;1; \boldsymbol{1}; \boldsymbol{1} \rangle_{0,0}  \oplus |1;0;0; \boldsymbol{1} ;\boldsymbol{1} \rangle_{0,0}  \Big ) }
       \\
       &\Big ( (0,0,0)_s \oplus (\tfrac12,\tfrac12,\tfrac12)_s \Big )) \otimes 
       \overline{\Big ( |1;0;0; \boldsymbol{248}; \boldsymbol{1} \rangle  \oplus |1;0;0;\boldsymbol{1};\boldsymbol{248} \rangle \Big )} \, ,
\end{aligned}
\end{equation}
where
\begin{equation}
    \begin{aligned}
        & (0,0,0)_s = |0;0;0\rangle \, ,
        \\
        & (\tfrac12,\tfrac12,\tfrac12)_s = |\tfrac12;\tfrac12;\tfrac12\rangle \oplus |1;1;0\rangle \oplus  |1;0;0\rangle \oplus  |1;0;1\rangle \oplus |\tfrac32;\tfrac12;\tfrac12\rangle \oplus |2;0;0\rangle \, .
    \end{aligned}
\end{equation}
The theory is supersymmetric, and indeed no unflowed continuous representations emerge.





\end{document}